\documentclass[aps,prc,twocolumn,superscriptaddress,showpacs,nofootinbib]{revtex4-1}
\usepackage{graphicx,amsmath,amssymb,bm,color}

\newcommand{\bra}[1]{\left\langle #1\right|}
\newcommand{\ket}[1]{\left| #1\right\rangle}
\newcommand{\kev}{\,\mathrm{keV}}
\newcommand{\mev}{\,\mathrm{MeV}}
\newcommand{\gev}{\,\mathrm{GeV}}

\newcommand{\fm}{\,\mathrm{fm}}

\newcommand{\beq}{\begin{equation}}
\newcommand{\eeq}{\end{equation}}
\newcommand{\nn}{\\ }
\newcommand{\fet}[1]{\mbox{\boldmath $#1$}}

\allowdisplaybreaks

\begin{document} 
 
\title{Local chiral effective field theory interactions
and quantum Monte Carlo applications} 
 
\author{A.\ Gezerlis}
\email[E-mail:~]{gezerlis@uoguelph.ca}
\affiliation{Department of Physics, 
University of Guelph, Guelph, Ontario N1G 2W1, Canada}
\author{I.\ Tews}
\email[E-mail:~]{tews@theorie.ikp.physik.tu-darmstadt.de}
\affiliation{Institut f\"ur Kernphysik,
Technische Universit\"at Darmstadt, 64289 Darmstadt, Germany}
\affiliation{ExtreMe Matter Institute EMMI,
GSI Helmholtzzentrum f\"ur Schwerionenforschung GmbH, 64291 Darmstadt, Germany}
\author{E.\ Epelbaum} 
\email[E-mail:~]{evgeny.epelbaum@ruhr-uni-bochum.de}
\affiliation{Institut f\"ur Theoretische Physik II, Ruhr-Universit\"at Bochum, 
44780 Bochum, Germany}
\author{M.\ Freunek} 
\affiliation{Institut f\"ur Theoretische Physik II, Ruhr-Universit\"at Bochum, 
44780 Bochum, Germany}
\author{S.\ Gandolfi}
\affiliation{Theoretical Division, Los Alamos National Laboratory,
Los Alamos, NM 87545, USA}
\author{K.\ Hebeler}
\affiliation{Institut f\"ur Kernphysik,
Technische Universit\"at Darmstadt, 64289 Darmstadt, Germany}
\affiliation{ExtreMe Matter Institute EMMI,
GSI Helmholtzzentrum f\"ur Schwerionenforschung GmbH, 64291 Darmstadt, Germany}
\author{A.\ Nogga}
\affiliation{Institut f\"ur Kernphysik, Institute for Advanced Simulation
and J\"ulich Center for Hadron Physics,
Forschungszentrum J\"ulich, 52425 J\"ulich, Germany}
\author{A.\ Schwenk}
\email[E-mail:~]{schwenk@physik.tu-darmstadt.de}
\affiliation{Institut f\"ur Kernphysik,
Technische Universit\"at Darmstadt, 64289 Darmstadt, Germany}
\affiliation{ExtreMe Matter Institute EMMI,
GSI Helmholtzzentrum f\"ur Schwerionenforschung GmbH, 64291 Darmstadt, Germany}

\begin{abstract} 

We present details of the derivation of local chiral effective field
theory interactions to next-to-next-to-leading order, and show results
for nucleon-nucleon phase shifts and deuteron properties for these
potentials. We then perform systematic auxiliary-field diffusion Monte
Carlo calculations for neutron matter based on the developed local
chiral potentials at different orders. This includes studies of the
effects of the spectral-function regularization and of the local regulators.
For all orders, we compare the quantum Monte Carlo results with
perturbative many-body calculations and find excellent agreement for
low cutoffs.

\end{abstract} 

\pacs{21.60.Ka, 21.30.-x, 21.65.Cd, 26.60.-c}
 
\maketitle 

\section{Introduction}

Chiral effective field theory (EFT) provides a systematic framework to
describe low-energy hadronic interactions based on the symmetries of
QCD.  In the past two decades, this method has been extensively
applied to nuclear forces and currents and to studies of the
properties of few- and many-nucleon systems, see
Refs.~\cite{Epelbaum:2009a,Entem:2011} for recent review articles. In
particular, accurate nucleon-nucleon (NN) potentials at
next-to-next-to-next-to-leading order (N$^3$LO) in the chiral
expansion have been
constructed~\cite{EGMN2LO,Entem:2003ft}. Presently, the main focus is
on the investigation of three-nucleon (3N) forces, see
Refs.~\cite{fewbody,Hammer:2013}, and on applications from light to
medium-mass
nuclei~\cite{NCSM,nuclattice,SM,Robert,CCreview,IMSRG,Ca,Gorkov}.

The available versions of the chiral potentials employ nonlocal
regularizations in momentum space and nonlocal contact interactions so
that the resulting potentials are strongly nonlocal. This feature
makes them not suitable for certain ab initio few- and many-body
techniques such as the quantum Monte Carlo (QMC) family of methods. As
we showed in our recent Letter~\cite{Gezerlis:2013ipa}, it is possible
to construct equivalent, local chiral NN potentials up to
next-to-next-to-leading order (N$^2$LO) by choosing a suitable set of
short-range operators and employing a local regulator. These local
potentials can be used in continuum QMC simulations because the
many-body propagator can be easily sampled.

The standard QMC approach used in the study of light nuclei
properties~\cite{Pudliner1997}, including
scattering~\cite{Nollett2007}, is the nuclear Green's Function Monte
Carlo (GFMC) method, which in addition to a stochastic integration
over the particle coordinates also performs explicit summations in
spin-isospin space~\cite{Carlson1987, Pieper:2001}. As a result, the
method is very accurate but computationally very costly and allows one
to access only nuclei with $A \leqslant 12$~\cite{Pieper2005,%
Lovato2013}. Larger particle numbers can be accessed with
Auxiliary-Field Diffusion Monte Carlo (AFDMC), which in addition to
the stochastic approach to the particle coordinates also
stochastically evaluates the summations in spin-isospin
space~\cite{Schmidt1999}, however at the cost of using simpler
variational wave functions than those used in nuclear GFMC. A new
Fock-space QMC method has recently been proposed in
Ref.~\cite{Roggero:2014}, which was used for a soft nonlocal
potential for pure neutron matter. In addition, an auxiliary-field QMC
study was recently carried out for a sharp-cutoff chiral
potential~\cite{Wlazlowski:2014jna}.

In this paper, we provide details of the derivation of local chiral
potentials to N$^2$LO and present tables of low-energy constants
(LECs) thus fully specifying the potential for use by others. We also
show results for phase shifts and deuteron properties. We then use the
new local chiral potentials in AFDMC simulations of neutron matter,
updating and augmenting our results of Ref.~\cite{Gezerlis:2013ipa},
and compare these to many-body perturbation theory (MBPT)
calculations.

\section{Local chiral potentials}

In chiral EFT, the different contributions to nuclear forces are
arranged according to their importance by employing a power-counting
scheme, see Refs.~\cite{Epelbaum:2009a,Entem:2011} and references
therein for more details. The NN potential is then given as a series
of terms
\begin{equation}
V_{\text{chiral}} = V^{(0)} + V^{(2)} + V^{(3)} + \dots \,,
\label{eq:ch_pot}
\end{equation}
where the superscript denotes the power in the expansion parameter $Q/
\Lambda_\text{b}$ with $Q$ referring to the soft scale associated with
typical momenta of the nucleons or the pion mass and $\Lambda_\text{b}
\sim M_\rho$ the hard scale corresponding to momenta at which the
chiral EFT expansion is expected to break down. We will take into
account all terms up to N$^2$LO in the chiral expansion. Generally,
one has to distinguish between two different types of contributions:
the long- and intermediate-range ones due to exchange of one or
several pions and the contact interactions, which parametrize the
short-range physics and are determined by a set of LECs fit to
experimental data. The long-range contributions are
completely determined by the chiral symmetry of QCD and low-energy
experimental data for the pion-nucleon system.

The crucial feature that allows us to construct a local version of the
chiral NN potential is the observation that the expressions for the
pion exchanges up to N$^2$LO only depend on the momentum transfer
${\bf q}={\bf p}'-{\bf p}$ with the incoming and outgoing relative
momenta ${\bf p}=({\bf p}_1-{\bf p}_2)/2$ and ${\bf p}'=({\bf
  p}'_1-{\bf p}'_2)/2$, respectively, provided the nucleon mass is
counted according to $Q/m_N \sim Q^2/\Lambda_\text{b}^2$ as suggested
in Ref.~\cite{Weinberg:1991um}.  Here, the ${\bf p}_i$ and ${\bf
  p}_i'$ correspond to incoming and outgoing momenta. This counting
scheme has been used in the derivation of nuclear
forces~\cite{EGMN2LO,Bernard:2011zr,Krebs:2012yv,Krebs:2013kha} and
electromagnetic currents~\cite{Kolling:2009iq,Kolling:2011mt} and has
as a consequence that the leading relativistic corrections to the
one-pion-exchange (OPE) potential enter at N$^3$LO.  Given that the
long-range potentials depend only on the momentum transfer, the
corresponding coordinate-space potentials are local. Here and in what
follows, we employ the decomposition for the long- and
intermediate-range potentials as
\begin{align}
V_{\text{long}} (r) &= V_C (r) + W_C (r) \; \fet \tau_1 \cdot \fet \tau_2 
\nonumber \\
& \quad + \bigl( V_S (r) + W_S (r) \; \fet \tau_1 \cdot \fet \tau_2 \bigr)
\, \fet \sigma_1 \cdot \fet \sigma_2 \nonumber \\
&\quad + \bigl( V_T (r) + W_T (r) \; \fet \tau_1 \cdot
\fet \tau_2 \bigr) S_{12}\,,
\end{align}
where ${\bf r}={\bf r}_1-{\bf r}_2$ denotes the separation between the
nucleons and $S_{12} = (3 \fet \sigma_1 \cdot \hat {\bf{r}} \; \fet
\sigma_2 \cdot \hat {\bf r} - \fet \sigma_1 \cdot \fet \sigma_2 )$ is
the tensor operator. The expression for the OPE potential at LO takes
the well-known form
\begin{align}
W_S^{(0)} (r ) &= \frac{M_{\pi}^3}{12 \pi} \left(\frac{g_A}{2
F_{\pi}}\right)^2 \frac{e^{-M_{\pi} r}}{M_{\pi} r} \,, \label{OPE_LO1} \nn
W_T^{(0)} (r ) &= \frac{M_{\pi}^3}{12 \pi} \left(\frac{g_A}{2
F_{\pi}}\right)^2 \frac{e^{-M_{\pi} r}}{M_{\pi} r} \, 
\left(1+\frac{3}{M_{\pi} r}+\frac{3}{(M_{\pi} r)^2} \right) \,, \label{OPE_LO2}
\end{align}
where $g_A$, $F_\pi$, and $M_\pi$ denote the axial-vector coupling
constant of the nucleon, the pion decay constant, and the pion mass,
respectively. In addition to these long-range contributions, the OPE
potential also involves a short-range piece proportional to a $\delta$
function. We absorb this contribution into the leading contact
interaction.

At next-to-leading order (NLO), the strength of the OPE potential is
slightly shifted due to the Goldberger-Treiman discrepancy
(GTD)~\cite{Fettes:1998ud}:
\begin{equation}
g_{\pi N}=\frac{g_{A} m_N}{F_{\pi}}\left(1-\frac{2M_{\pi}^{2}\,\bar{d}_{18}}{g_{A}}\right)\,,
\end{equation}
where $g_{\pi N}$ is the pion-nucleon coupling constant and
$\bar{d}_{18}$ is a LEC from the third-order pion-nucleon effective
Lagrangian, which is of the same order in the chiral expansion as
$V_{NN}^{(2)}$.

For the two-pion exchange (TPE) we use the
spectral-function-regularization (SFR) expressions as detailed in
Ref.~\cite{SF}. The coordinate-space expressions for the TPE potential
can be most easily obtained utilizing the spectral-function
representation with spectral functions $\rho_i$ and $\eta_i$:
\begin{align}
V_{C} (r) &= \frac{1}{2 \pi^2 r} \int_{2 M_\pi}^{\tilde \Lambda} d \mu \, \mu \, 
e^{-\mu r} \, \rho_{C} (\mu) \,, \label{four11} \nn
V_{S} (r) &= -\frac{1}{6 \pi^2 r} \int_{2 M_\pi}^{\tilde \Lambda} d \mu \, \mu \, 
e^{-\mu r} \, \Bigl( \mu^2 \rho_T (\mu) - 3 \rho_S (\mu ) \Bigr)\, , \label{four12} \nn
V_{T} (r) &= -\frac{1}{6 \pi^2 r^3} \int_{2 M_\pi}^{\tilde \Lambda}  d \mu \, \mu \, 
e^{-\mu r} \, \rho_T (\mu) \, ( 3 + 3 \mu r + \mu^2 r^2 )\,, 
\label{four13}
\end{align}
and similarly for $W_{C}$, $W_{S}$, and $W_{T}$ in terms of
$\eta_{C}$, $\eta_S$, and $\eta_T$ (instead of $\rho_{C}$, $\rho_S$,
and $\rho_T$).

In the framework of the SFR, the integrals in the spectral
representation of the TPE potential go from $2 M_\pi$ to the
ultraviolet cutoff $\tilde \Lambda$ rather than to $\infty$
corresponding to the case of dimensional regularization. Taking
$\tilde \Lambda$ of the order of $\Lambda_{\rm b}$
 ensures that no unnaturally large
short-range terms are induced by the subleading TPE
potential~\cite{SF}.

The TPE spectral functions at NLO are given by~\cite{DR} 
\begin{align}
\rho_{T}^{(2)} (\mu) &= \frac{1}{\mu^2}\, \rho_{S}^{(2)} (\mu) 
= \frac{3 g_A^4}{128 \pi F_\pi^4} \,
\frac{\sqrt{\mu^2 - 4 M_\pi^2}}{\mu} \label{spectr_nlo1}
\,, \nn
\eta_{C}^{(2)} (\mu) &= 
\frac{ 1}{768 \pi F_\pi^4}\,
\frac{\sqrt{\mu^2 - 4 M_\pi^2}}{\mu} \biggl(4M_\pi^2 (5g_A^4 - 4g_A^2 - 1) \nonumber \nn
& \quad- \mu^2(23g_A^4 - 10g_A^2 -1) 
+ \frac{48 g_A^4 M_\pi^4}{4 M_\pi^2 - \mu^2} \biggr) \,,\label{spectr_nlo2}
\end{align}
while the ones at N$^2$LO read 
\begin{align}
\rho_{C}^{(3)} (\mu) &= -\frac{3 g_A^2}{64 \mu F_\pi^4} \,(2M_\pi^2 - \mu^2) \, \Big(2M_\pi^2(2c_1 -c_3) + c_3 \mu^2 \Big) 
\,, \label{spectr_nnlo1}\nn 
\eta_{T}^{(3)} (\mu) &= \frac{1}{\mu^2}\, \eta_{S}^{(3)} (\mu) =
- \frac{g_A^2}{128 \mu F_\pi^4} \, c_4 (4M_\pi^2 - \mu^2)
\,,\label{spectr_nnlo2}
\end{align}
where $c_i$ denote the LECs of the subleading pion-nucleon
vertices~\cite{Bernard:1995dp}. For the subleading TPE potential, the
integrals in Eqs.~(\ref{four11})--(\ref{four13}) can be carried out
analytically leading to
\begin{align}
\label{coord_nnlo1}
V_C^{(3)} (r) &=  \frac{3 g_A^2}{32 \pi^2 F_\pi^4} \, \frac{e^{- 2 x}}{r^6}
\bigg[ 2 c_1 \, x^2 (1 + x)^2
\nonumber \\ & \quad + c_3 (6 + 12x + 10 x^2 + 4 x^3 + x^4) \bigg] \nonumber \nn
& \quad- {} \frac{3 g_A^2}{128 \pi^2 F_\pi^4} \, \frac{e^{- y}}{r^6} \bigg[
4 c_1 x^2 \Big(2 + y ( 2 + y) - 2 x^2   \Big) \nonumber \\
&\quad + {} c_3 \Big(  24 +  y (24 + 12 y  + 4 y^2 + y^3   )\nonumber \\
&\quad - 4 x^2 (2 + 2 y  + y^2 ) + 4  x^4 \Big) \bigg]\,, \\
W_S^{(3)} (r) &=  \frac{g_A^2}{48 \pi^2 F_\pi^4} \, \frac{e^{- 2 x}}{r^6}
c_4 \, (1 + x) (3 + 3 x + 2 x^2)\nonumber  \\
&\quad - {}  \frac{g_A^2}{384 \pi^2 F_\pi^4} \, \frac{e^{- y}}{r^6}
c_4 \, \Big( 24 + 24 y + 12 y^2 + 4 y^3 + y^4 
\nonumber \\ & \quad - 4 x^2 ( 2 + 2 y + y^2) \Big)\, ,\\
W_T^{(3)} (r) &= - \frac{g_A^2}{48 \pi^2 F_\pi^4} \, \frac{e^{- 2 x}}{r^6}
c_4 \, (1 + x) (3 + 3 x + x^2)  \nonumber\\
& \quad + {}  \frac{g_A^2}{768 \pi^2 F_\pi^4} \, \frac{e^{- y}}{r^6}
c_4 \, \Big( 48 + 48 y + 24 y^2 + 7 y^3 + y^4 
\nonumber \\ & \quad - 4 x^2 ( 8 + 5 y + y^2) \Big)\,,\label{coord_nnlo3}
\end{align}
where we have introduced dimensionless variables $x \equiv M_\pi r$ and
$y \equiv \tilde \Lambda r$. Analytic expressions for the leading TPE potentials for
the case of $\tilde \Lambda = \infty$ are given in Ref.~\cite{DR}. 

We now turn to the short-range contact interactions, starting from the
expressions in momentum space.  The most general set of contact
interactions at LO is given by momentum-independent terms $\openone,
\, {\bm \sigma}_1 \cdot {\bm \sigma}_2, \, {\bm \tau}_1 \cdot {\bm
  \tau}_2$, and ${\bm \sigma}_1 \cdot {\bm \sigma}_2 \, {\bm \tau}_1
\cdot {\bm \tau}_2$, so that one has
\begin{equation}
V_ {\text{cont}}^{(0)} = \alpha_1 + \alpha_2 {\bm \sigma}_1 \cdot {\bm \sigma}_2 + \alpha_3 {\bm \tau}_1 \cdot {\bm \tau}_2 + \alpha_4 {\bm \sigma}_1 \cdot {\bm \sigma}_2 \, {\bm \tau}_1 \cdot {\bm \tau}_2 \,.
\label{eq:ch_cont_LO}
\end{equation}
As discussed below, out of these four terms only two are linearly
independent.  As nucleons are fermions, they obey the Pauli principle,
and after antisymmetrization the potential $V$ is given by:
\begin{equation}
V_{\text{A}} = \frac12 \left(V-\mathcal{A}[V] \right),
\label{eq:antisym_pot}
\end{equation}
with the exchange operator $\mathcal{A}$ defined as
\begin{align}
\mathcal{A}[V({\bf q},{\bf k})]&=\frac14 (1+{\bm \sigma}_1 \cdot {\bm
\sigma}_2)(1+{\bm \tau}_1 \cdot {\bm \tau}_2) \nonumber \\
& \quad \times V\left({\bf q} \rightarrow -2{\bf k}, {\bf k} \rightarrow -\frac12 {\bf q}\right),
\label{eq:antisymmetrizer}
\end{align}
where  ${\bf k}=({\bf p}'+{\bf p})/2$ is the momentum transfer in the exchange channel.  
For the LO contact potential, we have
\begin{align}
V^{(0)}_{\text{cont,A}} &= \frac12 \left(1- \frac14 (1+{\bm \sigma}_1 \cdot {\bm \sigma}_2)(1+{\bm \tau}_1 \cdot {\bm \tau}_2)\right) V_ {\text{cont}}^{(0)} \nonumber \\
&= \left(\frac{3}{8}\alpha_1 -\frac{3}{8}\alpha_2 -\frac{3}{8}\alpha_3 -\frac{9}{8}\alpha_4 \right) \nonumber \\
& \quad + \left(-\frac{1}{8}\alpha_1 +\frac{5}{8}\alpha_2 -\frac{3}{8}\alpha_3 +\frac{3}{8}\alpha_4 \right) {\bm \sigma}_1 \cdot {\bm \sigma}_2 \nonumber \\
& \quad + \left(-\frac{1}{8}\alpha_1 -\frac{3}{8}\alpha_2 +\frac{5}{8}\alpha_3 +\frac{3}{8}\alpha_4 \right) {\bm \tau}_1 \cdot {\bm \tau}_2 \nonumber \\
& \quad + \left(-\frac{1}{8}\alpha_1 +\frac{1}{8}\alpha_2 +\frac{1}{8}\alpha_3 +\frac{3}{8}\alpha_4 \right) {\bm \sigma}_1 \cdot {\bm \sigma}_2 \, {\bm \tau}_1 \cdot {\bm \tau}_2 \nonumber \\
&=\tilde C_S + \tilde C_T {\bm \sigma}_1 \cdot {\bm \sigma}_2 + \left(-\frac{2}{3}\tilde C_S-\tilde C_T \right) {\bm \tau}_1 \cdot {\bm \tau}_2 \nonumber \\
&\quad + \left(-\frac13 \tilde C_S\right) {\bm \sigma}_1 \cdot {\bm \sigma}_2 \, {\bm \tau}_1 \cdot {\bm \tau}_2 \,.
\label{eq:LO_antisymm}
\end{align}
Obviously, there are only two independent couplings at leading order
after antisymmetrization. Following Weinberg~\cite{Weinberg:1991um},
we take
\begin{equation}
V^{(0)}_{\text{cont}} = C_S + C_T {\bm \sigma}_1 \cdot {\bm \sigma}_2\,,
\label{eq:antisym_LO}
\end{equation}
but we could have chosen different two of the four contact
interactions. This is analogous to Fierz ambiguities.  At NLO, 14
different contact interactions are allowed by symmetries:
\begin{align}
V_ {\rm cont}^{(2)} &= \gamma_1 \, q^2 + \gamma_2 \, q^2\, {\bm \sigma}_1 \cdot {\bm \sigma}_2 + \gamma_3 \, q^2\, {\bm \tau}_1 \cdot {\bm \tau}_2 \nonumber \\
&\quad + \gamma_4 \, q^2 {\bm \sigma}_1 \cdot {\bm \sigma}_2 {\bm \tau}_1 \cdot {\bm \tau}_2 \nonumber \\
&\quad + \gamma_5 \, k^2 + \gamma_6 \, k^2\, {\bm \sigma}_1 \cdot {\bm \sigma}_2 + \gamma_7 \, k^2\, {\bm \tau}_1 \cdot {\bm \tau}_2 \nonumber \\
&\quad + \gamma_8 \, k^2 {\bm \sigma}_1 \cdot {\bm \sigma}_2 {\bm \tau}_1 \cdot {\bm \tau}_2 \nonumber \\
&\quad + \gamma_9 \, ({\bm \sigma}_1 + {\bm \sigma}_2)({\bf q}\times {\bf k})\nonumber \\ 
&\quad + \gamma_{10} \, ({\bm \sigma}_1 + {\bm \sigma}_2)({\bf q}\times {\bf k}) {\bm \tau}_1 \cdot {\bm \tau}_2 \nonumber \\
&\quad + \gamma_{11} ({\bm \sigma}_1 \cdot {\bf q}) ({\bm \sigma}_2 \cdot {\bf q}) \nonumber \\
&\quad + \gamma_{12} ({\bm \sigma}_1 \cdot {\bf q}) ({\bm \sigma}_2 \cdot {\bf q}) {\bm \tau}_1 \cdot {\bm \tau}_2 \nonumber \\
&\quad + \gamma_{13} ({\bm \sigma}_1 \cdot {\bf k}) ({\bm \sigma}_2 \cdot {\bf k}) \nonumber \\
&\quad + \gamma_{14} ({\bm \sigma}_1 \cdot {\bf k}) ({\bm \sigma}_2 \cdot {\bf k}) {\bm \tau}_1 \cdot {\bm \tau}_2 \,.
\label{eq:NLOlong}
\end{align}
In analogy to the LO case, only seven couplings are independent and
one has the freedom to choose an appropriate basis. The currently
available versions of chiral potentials~\cite{EGMN2LO,Entem:2003ft}
use the set which does not involve isospin operators.  Because we want
to construct a local chiral potential, we have to eliminate contact
interactions that depend on the momentum transfer in the exchange
channel ${\bf k}$. Thus, we choose
\begin{align} 
V^{(2)}_{\rm cont} &= C_1 \, q^2 + C_2 \, q^2 \, 
{\bm \tau}_1 \cdot {\bm \tau}_2 \nonumber \\
&\quad  + \bigl(C_3 \, q^2 + C_4 \, q^2 \, {\bm \tau}_1 \cdot {\bm \tau}_2 \bigr)
\, {\bm \sigma}_1 \cdot {\bm \sigma}_2 \nonumber \\
&\quad + i \, \frac{C_5}{2} \, ({\bm \sigma}_1 + {\bm \sigma}_2) \cdot
{\bf q} \times {\bf k} \nonumber \\
&\quad + C_6 \, ({\bm \sigma}_1 \cdot {\bf q})({\bm \sigma}_2 \cdot {\bf q}) 
\nonumber \\
&\quad + C_7 \, ({\bm \sigma}_1 \cdot {\bf q})({\bm \sigma}_2 \cdot {\bf q}) 
\, {\bm \tau}_1 \cdot {\bm \tau}_2 \,,
\label{eq:NLOshort}
\end{align}
which is local except for the ${\bf k}$-dependent spin-orbit
interaction $(C_5)$. Without regulators, the expressions for the
contact interactions in coordinate space are of the form
\begin{align}
V^{(0)}_{\rm cont}({\bf r}) &=  ( C_S   + C_T {\bm \sigma}_1 \cdot {\bm
\sigma}_2 ) \delta ({\bf r}) \,,\label{eq:LO_FT}\nn
V^{(2)}_{\rm cont}({\bf r}) &= -(C_1+C_2 {\bm \tau}_1 \cdot {\bm
\tau}_2)\Delta \delta ({\bf r}) \nonumber \\
&\quad -(C_3+C_4 {\bm \tau}_1 \cdot {\bm \tau}_2) \, {\bm \sigma}_1 \cdot
{\bm \sigma}_2 \Delta \delta ({\bf r})\nonumber \\
&\quad + \frac{C_5}{2}\frac{\partial_r \delta ({\bf r})}{r} {\bf L}\cdot {\bf S}+ (C_6+ C_7 {\bm \tau}_1 \cdot {\bm \tau}_2) \nonumber \\
&\quad \times \left[({\bm \sigma}_1 \cdot \hat{\bf r})({\bm \sigma}_2
\cdot \hat{\bf r}) \left(\frac{\partial_r \delta ({\bf
r})}{r}- \partial_r^2 \delta ({\bf r}) \right) \right. \nonumber \\
&\quad -\left. {\bm \sigma}_1 \cdot {\bm \sigma}_2 \frac{\partial_r \delta ({\bf r})}{r}  \right]\,.
\label{eq:NLO_FT}
\end{align}
The derivation of these expressions is given in Appendix~\ref{PWD}.

In addition to the isospin-symmetric contributions to the potential
given by Eqs.~(\ref{OPE_LO1}), (\ref{OPE_LO2}),
(\ref{four11})--(\ref{four13}),
(\ref{coord_nnlo1})--(\ref{coord_nnlo3}), (\ref{eq:LO_FT}), and
(\ref{eq:NLO_FT}), we take into account isospin-symmetry-breaking
corrections~\cite{Epelbaum:2005fd}. We include long-range
charge-independence breaking (CIB) terms due to the pion mass
splitting in the OPE potential,
\begin{align}
V_{\rm OPE, \; full} &= V_{\rm OPE} (M_{\pi^\pm}) \fet \tau_1 \cdot \fet \tau_2 \nonumber \nn 
&\quad + 4 \Big[ V_{\rm OPE} (M_{\pi^0}) -V_{\rm OPE} (M_{\pi^\pm}) \Big] \tau_1^3 \tau_2^3\,,
\end{align}
where $V_{\rm OPE}$ is given by
\begin{align}
V_{\rm OPE}(M) &= \frac{M^3}{12 \pi} \left(\frac{g_A}{2
F_{\pi}}\right)^2 \frac{e^{-M r}}{M r} \Bigg[\fet \sigma_1 \cdot \fet \sigma_2  \nonumber \\ & \quad \times 
\left(1+\frac{3}{M r}+\frac{3}{(M r)^2} \right)S_{12}\Bigg]\,. \label{eq:OPE_IB}
\end{align}
For the contact interactions, we include the leading
momentum-independent CIB and charge-symmetry-breaking (CSB) terms,
which in coordinate space have the form
\begin{align}
\label{cont_IB}
V_{\rm cont, \; CIB} ({\bf r}) &= C_{\rm CIB} \frac{1 + 4 \tau_1^3 \tau_2^3 }{2}
\,  \frac{1 - \fet \sigma_1 \cdot \fet \sigma_2 }{4} \delta ({\bf r}) \,, \nn
V_{\rm cont, \; CSB} ({\bf r}) &= C_{\rm CSB} (\tau_1^3 + \tau_2^3 )
\,  \frac{1 - \fet \sigma_1 \cdot \fet \sigma_2 }{4} \delta ({\bf r})\,.\label{cont_IB2}
 \end{align}
These contact interactions are defined in such a way that they do not
affect neutron-proton observables.  Furthermore, the last factor, $(1
- \fet \sigma_1 \cdot \fet \sigma_2)/4$, is a projection operator on
spin-0 states and ensures that spin-triplet partial waves are not
affected by the above terms. This factor is redundant for
non-regularized expressions. Note that the impact of the spin-0
projection on NN phase shifts is very small, typically between
$1-2\%$. This is smaller than the deviation from the phase shifts of
the Nijmegen partial wave analysis (PWA) and smaller than the theoretical
uncertainty of the results. Thus, in the following we will neglect the
spin-0 projection factor.

\begin{table*}[t]
\begin{center}
\caption{Low-energy constants for $R_0=1.0, 1.1, 1.2 \, {\rm fm}$ 
at LO, NLO, and N$^2$LO (with a spectral-function cutoff
$\tilde{\Lambda}=1000 \, {\rm MeV}$). The couplings $C_{1-7}$ are
given in fm$^4$ while the rest are in fm$^2$.\label{tab:contacts10}}
\begin{tabular}{l|ccc|ccc|ccc}
\hline
$R_0$ &  & 1.0 fm &  &  & 1.1 fm & &  & 1.2 fm & \\
& LO & NLO & N$^2$LO & LO & NLO & N$^2$LO& LO & NLO & N$^2$LO\\
\hline
$C_S$ & $-0.75112$ & \,\,\,\,\,$3.16803$ & \,\,\,\,\,$5.43850$ & $-1.29631$ & \,\,\,\, $1.03075$ & \,\,\,\,\,$3.88699$ & $-1.79693$ & \,\,\,\, $0.03551$ & \,\,\,\,\,$2.68765$ \\
$C_T$ & \,\,\,\,\,$0.37409$ & \,\,\,\, $1.41396$ & \,\,\,\,$ 0.27672$ & \,\,\,\,\,$0.25648$ & \,\,\,\,\,$0.90699$ & \,\,\,\,$0.24416$& \,\,\,\,\,$0.15442$ & \,\,\,\,\,$0.71729$ & \,\,\,\, $0.23382$\\
$C_1$ & & \,\,\,\,\,$0.31420$ & $-0.14084$ & & \,\,\,\,\,$0.27239$ & $-0.09650$& & \,\,\,\,\,$0.22288$ & $-0.07951$\\
$C_2$ & & \,\,\,\,\,$0.25786$ & \,\,\,\, $0.04243$ & & \,\,\,\,\,$0.22032$ & \,\,\,\, $0.05947$& & \,\,\,\,\,$0.22878$ & \,\,\,\, $0.07610$\\
$C_3$ & & $-0.13134$ & $-0.12338$ & & $-0.13641$ & $-0.14183$& & $-0.15043$ & $-0.16926$ \\
$C_4$ & & \,\,\,\, $0.11861$ & \,\,\,\, $0.11018$ & & \,\,\,\, $0.09420$ & \,\,\,\,\,$0.11146$& & \,\,\,\, $0.08929$ & \,\,\,\,\,$0.12359$\\
$C_5$ & & \,\,\,\, $2.38552$ & \,\,\,\, $2.11254$  & & \,\,\,\, $2.16238$ & \,\,\,\, $2.0082$  & & \,\,\,\, $2.02932$ & \,\,\,\, $1.94280$ \\
$C_6$ & & \,\,\,\,\,$0.37319$ & \,\,\,\,\,$0.15898$ & & \,\,\,\,\,$0.33065$ & \,\,\,\,\,$0.18318$& & \,\,\,\,\,$0.34011$ & \,\,\,\,\,$0.21421$\\
$C_7$ & & $-0.35668$ & $-0.26994$ & & $-0.33570$ & $-0.30105$& & $-0.36248$ & $-0.34193$\\
$C_{\text{CIB}}$ & $-0.02361$ & \,\,\,\,$0.05094$ & \,\,\,\,\,$0.05320$ & $-0.01922$ & \,\,\,\, $0.05153$ & \,\,\,\,\,$0.05538$& $-0.01335$ & \,\,\,\, $0.05477$ & \,\,\,\,\,$0.05648$\\
$C_{\text{CSB}}$ & $-0.01988$ & \,\,\,\,$0.00823$ & \,\,\,\,\,$0.00976$ & $-0.02001$ & \,\,\,\, $0.00704$ & \,\,\,\,\,$0.00902$ & $-0.01959$ & \,\,\,\, $0.00660$ & \,\,\,\,\,$0.00771$\\
\hline 
\end{tabular}
\end{center}
\end{table*}

\begin{table*}[t]
\begin{center}
\caption{Low-energy constants for $R_0=1.0, 1.1, 1.2 \, {\rm fm}$ 
at LO, NLO, and N$^2$LO (with a spectral-function cutoff
$\tilde{\Lambda}=1400 \mev$). The couplings $C_{1-7}$ are
given in fm$^4$ while the rest are in fm$^2$.\label{tab:contacts14}}
\begin{tabular}{l|ccc|ccc|ccc}
\hline
$R_0$ &  & 1.0 fm &  &  & 1.1 fm & &  & 1.2 fm & \\
& LO & NLO & N$^2$LO & LO & NLO & N$^2$LO& LO & NLO & N$^2$LO\\
\hline
$C_S$ & $-0.75112$ & \,\,\,\,\,$3.32404$ & \,\,\,\,\,$8.16454$ & $-1.29631$ & \,\,\,\, $1.13903$ & \,\,\,\,\,$5.89685$& $-1.79693$ & \,\,\,\, $0.10909$ & \,\,\,\,\,$4.19629$ \\
$C_T$ & \,\,\,\,\,$0.37409$ & \,\,\,\, $1.30221$ & $-0.14809$ & \,\,\,\,\,$0.25648$ & \,\,\,\,\,$0.81867$ & $-0.08689$& \,\,\,\,\,$0.15442$ & \,\,\,\,\,$0.64646$ & $-0.02820$\\
$C_1$ & & \,\,\,\,\,$0.30649$ & $-0.12250$ & & \,\,\,\,\,$0.25830$ & $-0.04061$& & \,\,\,\,\,$0.21280$ & \,\,\,\, $0.00211$\\
$C_2$ & & \,\,\,\,\,$0.26558$ & \,\,\,\, $0.00843$ & & \,\,\,\,\,$0.23565$ & \,\,\,\, $0.02161$& & \,\,\,\,\,$0.24032$ & \,\,\,\, $0.03805$\\
$C_3$ & & $-0.14378$ & $-0.12964$ & & $-0.14535$ & $-0.15446$& & $-0.16477$ & $-0.18525$ \\
$C_4$ & & \,\,\,\, $0.13434$ & \,\,\,\, $0.12390$ & & \,\,\,\, $0.10401$ & \,\,\,\,\,$0.12110$& & \,\,\,\, $0.10228$ & \,\,\,\,\,$0.12819$\\
$C_5$ & & \,\,\,\, $2.39094$ & \,\,\,\, $2.13434$  & & \,\,\,\, $2.16525$ & \,\,\,\, $2.02482$  & & \,\,\,\, $2.02827$ & \,\,\,\, $1.95804$ \\
$C_6$ & & \,\,\,\,\,$0.38680$ & \,\,\,\,\,$0.12495$ & & \,\,\,\,\,$0.34394$ & \,\,\,\,\,$0.14992$& & \,\,\,\,\,$0.35219$ & \,\,\,\,\,$0.18335$\\
$C_7$ & & $-0.37920$ & $-0.27533$ & & $-0.35731$ & $-0.30346$& & $-0.38191$ & $-0.34227$\\
$C_{\text{CIB}}$ & $-0.02361$ & \,\,\,\,$0.05088$ & \,\,\,\,\,$0.05290$ & $-0.01922$ & \,\,\,\, $0.05151$ & \,\,\,\,\,$0.05538$& $-0.01335$ & \,\,\,\, $0.05468$ & \,\,\,\,\,$0.05592$\\
$C_{\text{CSB}}$ & $-0.01988$ & \,\,\,\,$0.00821$ & \,\,\,\,\,$0.00961$ & $-0.02001$ & \,\,\,\, $0.00701$ & \,\,\,\,\,$0.00883$ & $-0.01959$ & \,\,\,\, $0.00652$ & \,\,\,\,\,$0.00714$\\
\hline 
\end{tabular}
\end{center}
\end{table*}

\begin{figure*}[t]
\centering
\includegraphics[height=0.245\textwidth]{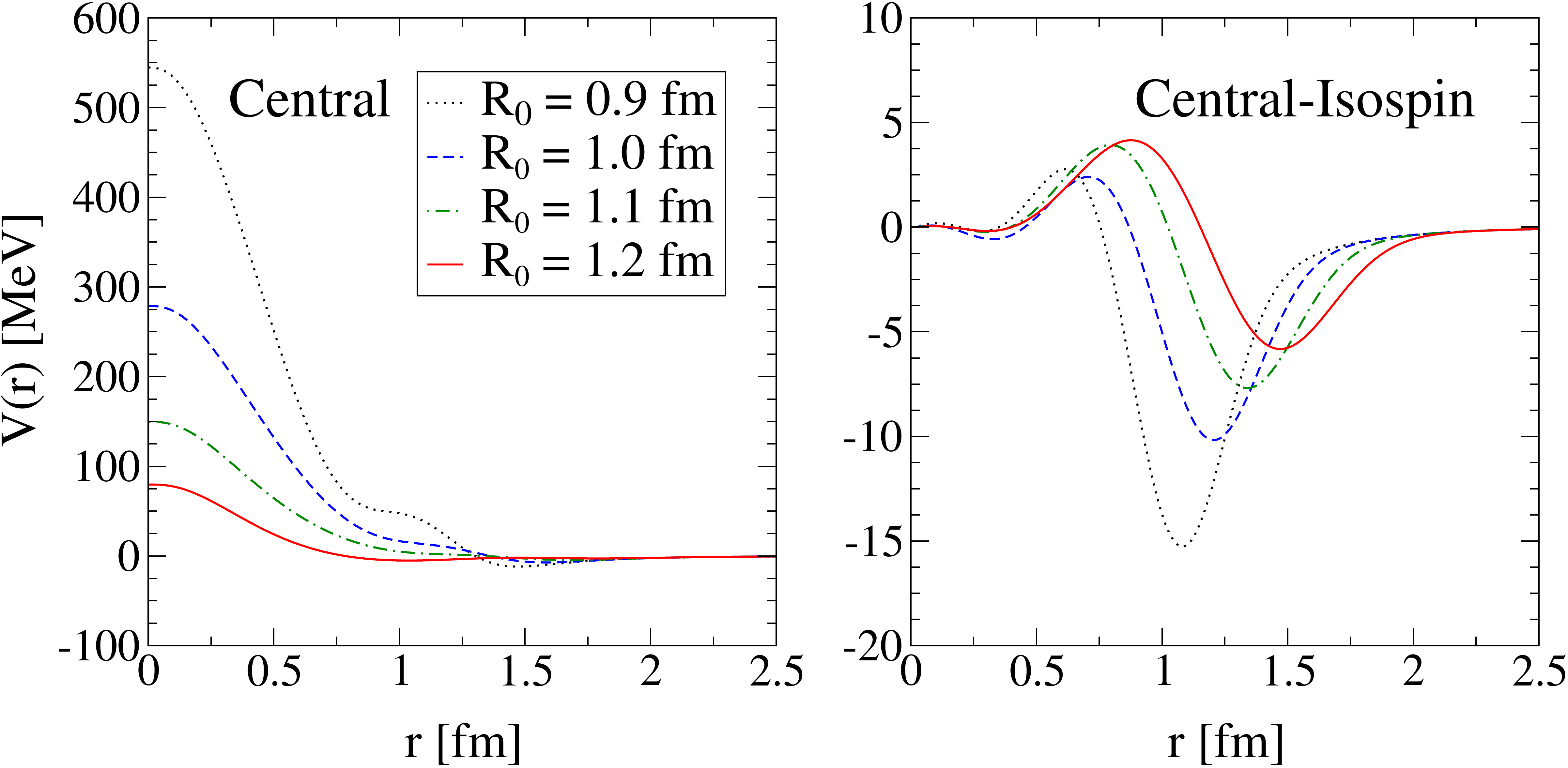} \hspace*{0.1cm}
\includegraphics[height=0.245\textwidth]{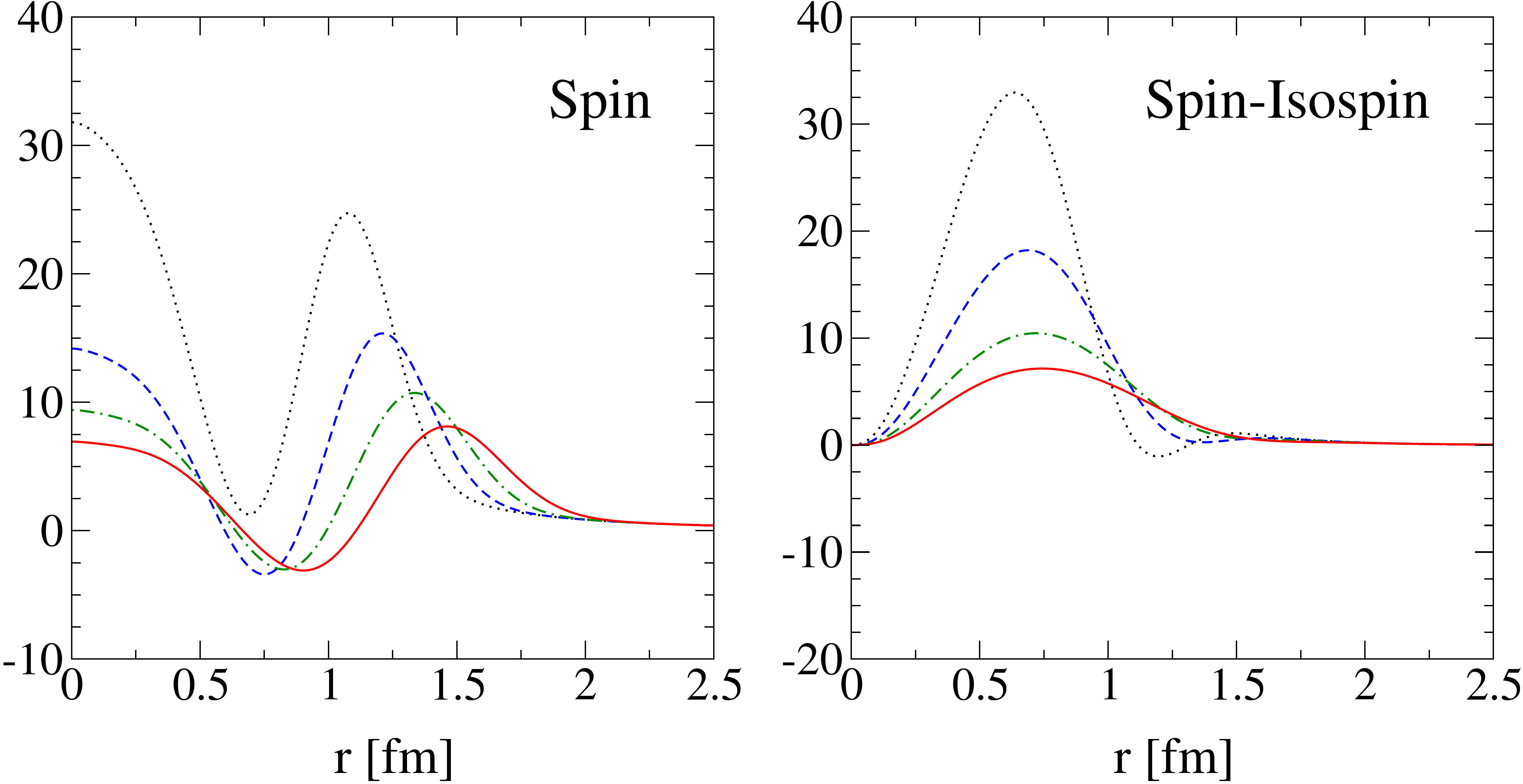} \\
\vspace{0.15cm}
\includegraphics[height=0.245\textwidth]{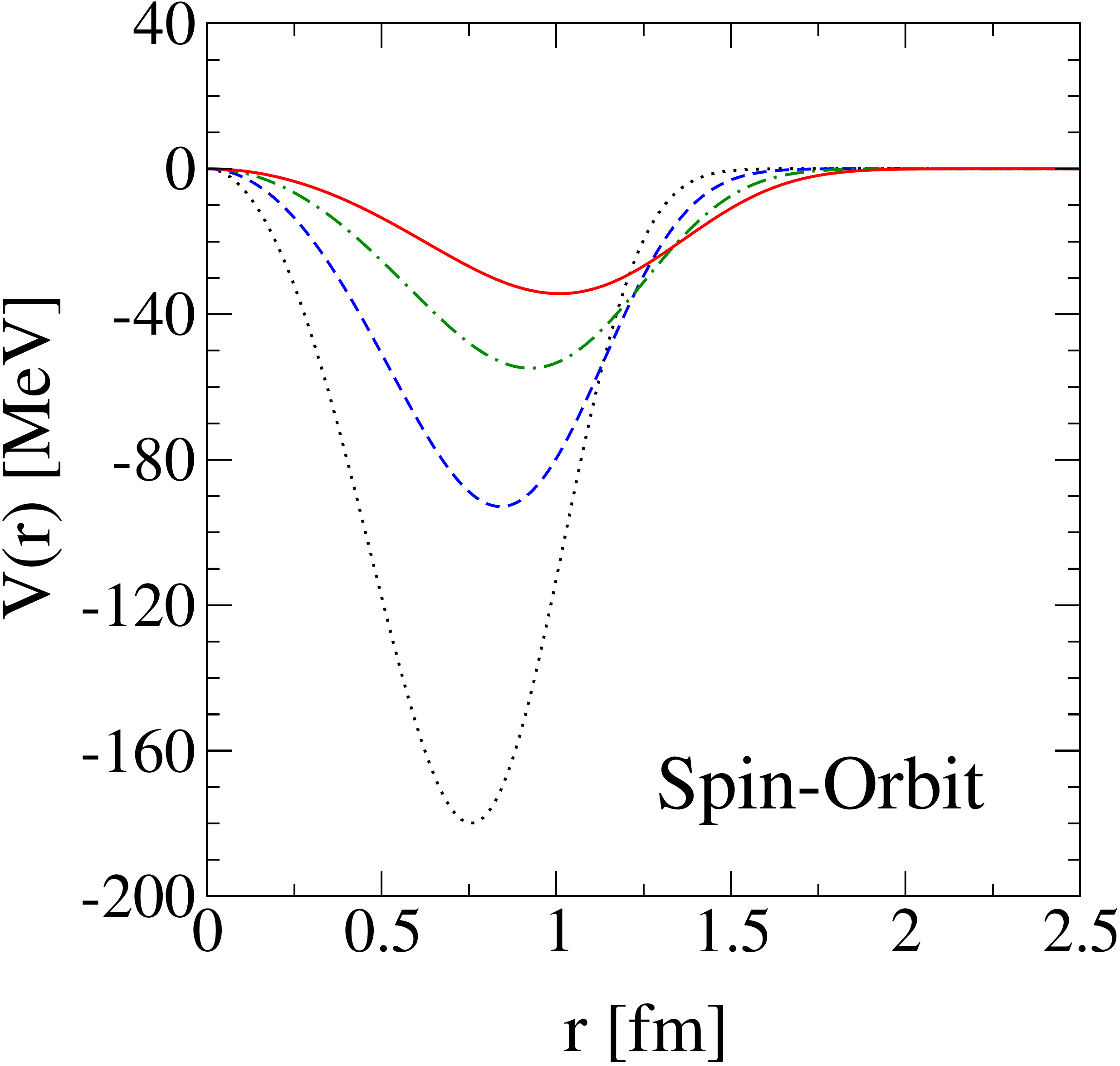}\hspace*{0.15cm}
\includegraphics[height=0.245\textwidth]{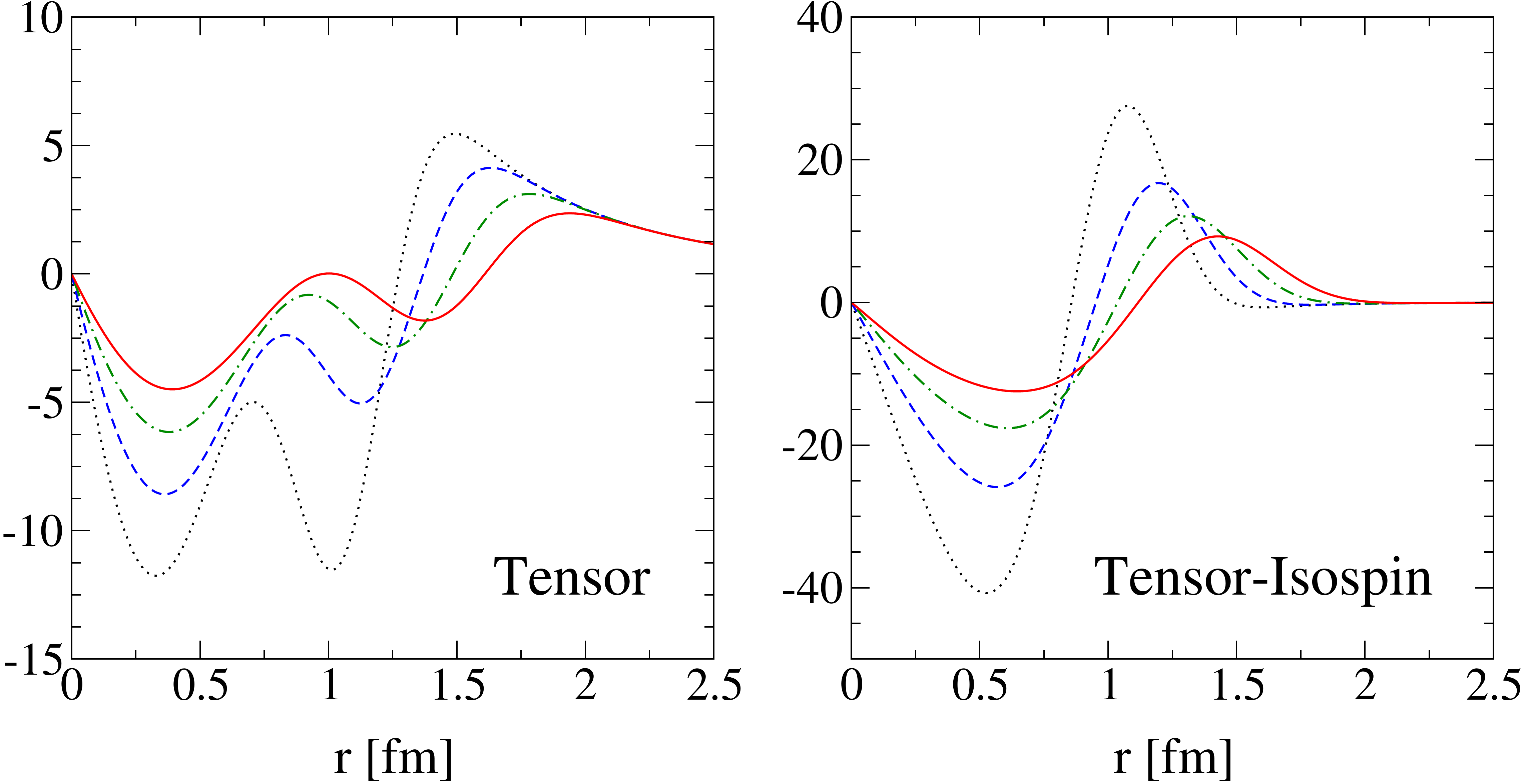}
\caption{(Color online) Local chiral NN potentials $V(r)$ at N$^2$LO
for an SFR cutoff $\tilde{\Lambda}=1000 \mev$, decomposed into the
central, central-isospin, spin, spin-isospin, spin-orbit, tensor, and
tensor-isospin components, for cutoffs $R_0=0.9-1.2 \fm$. For all
components, we observe a softening of the potential going from a
cutoff $R_0=0.9 \fm$ to $R_0=1.2\fm$. We include the $R_0=0.9 \fm$
potential for illustration, but as discussed in the text, will not
use it in many-body calculations.\label{fig:PotChannel}}
\end{figure*}

We are now in the position to specify the regularization scheme for
the NN potential. Following Ref.~\cite{Gezerlis:2013ipa}, this is
achieved by multiplying the coordinate-space expressions for the
long-range potential in Eqs.~(\ref{OPE_LO1}), (\ref{OPE_LO2}),
(\ref{four11})--(\ref{four13}), and
(\ref{coord_nnlo1})--(\ref{coord_nnlo3}) with a regulator function
\beq
V_{\rm long } (r) \; \to \;  V_{\rm long } (r)  \Big( 1 -
e^{-(r/R_0)^4} \Big)\,. 
\eeq
This ensures that short-distance parts of the long-range potentials at
$r$ smaller than $R_0$ are smoothly cut off.  For the short-range
terms in Eqs.~(\ref{eq:LO_FT}), (\ref{eq:NLO_FT}), (\ref{cont_IB}),
and (\ref{cont_IB2}) the regularization is achieved by employing a
local regulator $f_{\rm local}(q^2)$, leading to the replacement of
the $\delta$-function by a smeared one with the same exponential
smearing factor as for the long-range regulator,
\beq
\delta ({\bf r}) \; \to \; \delta_{R_0} ({\bf{r}}) = \alpha e^{-
(r/R_0)^4} \,,
\eeq 
where the normalization constant, 
\beq
\alpha = \frac{1}{\pi \Gamma\bigl(3/4\bigr) R_0^3}\, ,
\eeq
ensures that 
\beq
\int d^3 r \, \delta_{R_0} ({\bf{r}}) = 1\,. 
\eeq
The Fourier transformations of the contact interactions taking into
account the local regulator $f_{\rm local}(q^2)$ are given in
Appendix~\ref{Fouriertrafo}. The choice of the coordinate-space cutoff
$R_0$ is dictated by the following considerations. On the one hand,
one would like to take $R_0$ as small as possible to ensure that one
keeps most of the long-range physics associated with the pion-exchange
potentials. On the other hand, it is shown in Ref.~\cite{Baru:2012iv}
that at least for the particular class of pion-exchange diagrams
corresponding to the multiple-scattering series, the chiral expansion
for the NN potential breaks down at distances of the order of
$r\sim0.8 \fm$ but converges fast for $r \gtrsim 1 \fm$. This suggests
that a useful choice of the cutoff $R_0$ is $R_0 \sim 1 \fm$, which
corresponds to momentum-space cutoffs of the order of $\sim
500 \mev$. This follows from Fourier transforming the regulator
function, integrating it from $0$ to infinity, and comparing to a
sharp cutoff. These values are similar to the ones adopted in the
already existing, nonlocal implementations of the chiral
potential~\cite{EGMN2LO,Entem:2003ft}, see also
Ref.~\cite{Rentmeester:1999vw, Marji:2013uia} for a related
discussion.

In view of the arguments provided in
Refs.~\cite{Lepage:1997cs,Epelbaum:2009sd,Zeoli:2012bi}, we will not
use significantly lower values of $R_0$ in applications,~\footnote{See, however,
Ref.~\cite{Epelbaum:2012ua} where a new, renormalizable approach to
NN scattering is formulated that allows to completely eliminate the
ultraviolet cutoff.} although we were able to obtain
fits to NN phase shifts using $R_0 = 0.9 \fm$. However, the LECs start
to become unnatural for this cutoff. On the other hand, choosing
considerably larger values of $R_0$ results in cutting off the
long-range physics we want to preserve and, thus, introduces an
unnecessary limitation in the breakdown momentum of the
approach. Therefore, here and in the following, we will allow for a
variation of the cutoff $R_0$ in the range of $R_0 = 1.0 - 1.2 \fm$.
Because the local regulator eliminates a considerable part of
short-distance components of the TPE potential, we are much less
sensitive to the choice of the SFR cutoff $\tilde \Lambda$ as compared
to Refs.~\cite{EGMN2LO,Epelbaum:2003xx} and can safely increase it up
to $\tilde \Lambda = 1.4$ GeV without producing spurious deeply bound
states. In this work, we will vary $\tilde \Lambda$ in
the range $\tilde \Lambda = 1.0 - 1.4$ GeV. In future work, we will
explore removing the SFR cutoff $\tilde{\Lambda} \to \infty$.

We would like to underline that there is no conceptual difference
between our local regularization and the nonlocal regularization
currently used in widely employed versions of chiral interactions in
momentum space. The local chiral potentials include the same physics
as the momentum-space versions. This is especially clear when
antisymmetrizing. The local regulator by construction preserves the
long-range parts of the interaction. When Fourier transformed, it 
generates higher-order q$^2$-dependent terms when applied to 
short-range operators, like those already present at NLO and N$^2$LO. 
Note that antisymmetrization and
local regularization do not commute, but the commutator is given by
higher-order terms. At NLO and N$^2$LO, the $2+7$ contact interactions
provide a most general representation consistent with all symmetries.

It remains to specify the values of the LECs and masses that enter the
NN potentials at N$^2$LO.  In the following, we use $m_p=938.272
\mev$, $m_n=939.565 \mev$, the average pion mass $M_{\pi}=138.03
\mev$, the pion decay constant $F_{\pi}=92.4 \mev$, and the axial
coupling $g_A=1.267$.  For the pion-nucleon coupling, we adopt the
value of $g^2_{\pi N}/(4\pi )= 13.54$ which is consistent with
Ref.~\cite{Timmermans:1990tz}, which also agrees with the recent
determination in Ref.~\cite{Baru:2010xn} based on the
Goldberger-Miyazawa-Oehme sum rule and utilizing the most accurate
available data on the pion-nucleon scattering lengths.  In order to
account for the GTD as described above, we use the value $g_A = 1.29$
in the expressions for the OPE potential.  For the LECs $c_i$ in the
N$^2$LO TPE potential, we use the same values as in
Ref.~\cite{EGMN2LO}, namely $c_1=-0.81 \gev^{-1}, c_3=-3.4 \gev^{-1}$,
and $c_4=3.4 \gev^{-1}$.

We emphasize that we use the same expression for the OPE potential
that includes isospin-symmetry-breaking corrections and accounts for
the GTD as well as the same isospin-symmetry-breaking contact
interactions at all orders in the chiral expansion to allow for a more
meaningful comparison between LO, NLO and N$^2$LO.

With the NN potential specified as above, we have performed
$\chi^2$-fits to neutron-proton phase shifts from the Nijmegen
PWA~\cite{Stoks:1993tb} for $R_0 = 0.9$, $1.0$, $1.1$ and $1.2 \fm$
and $\tilde \Lambda = 0.8$, $1.0$, $1.2$ and $1.4$ GeV. We used the
separation of spin-singlet and spin-triplet channels, and, at LO, fit
the $^1S_0$ and $^3S_1$ partial waves separately while at NLO and
N$^2$LO we fit the $\{^1S_0, ^1P_1\}$ and $\{^3S_1, \epsilon_1, ^3P_0,
^3P_1, ^3P_2\}$ partial waves. At NLO and N$^2$LO, we used the same
energies of $E_{\rm lab} =1$, $5$, $10$, $25$, $50$, $100$ and $150
\mev$ for $R_0 = 1.0$ and $R_0 = 1.1 \fm$ as in the Nijmegen PWA and
the errors in the phase shifts provided in
Ref.~\cite{Stoks:1993tb}. For $R_0 = 1.2 \fm$, the fits are performed
up to $E_{\rm lab}= 100 \mev$. At LO, the fits are performed up to
$E_{\rm lab} = 50 \mev$. Finally, the values of the LECs $C_{\rm CIB}$
and $C_{\rm CSB}$ are adjusted to reproduce the proton-proton $^1S_0$
scattering length $a_{pp} = -7.81 \fm$ and the recommended value of
the neutron-neutron scattering length $a_{nn} = -18.9 \fm$.  Note that
we only take into account the point-like Coulomb force for the
electromagnetic interaction as appropriate to N$^2$LO, see
Ref.~\cite{EGMN2LO} for more details.  The resulting LECs for
$R_0=1.0, 1.1, 1.2 \fm$ and $\tilde{\Lambda}=1000 \mev$ are shown in
Table \ref{tab:contacts10} and for $\tilde{\Lambda}=1400 \mev$ in
Table \ref{tab:contacts14}.

\begin{figure}[t]
\centering
\includegraphics[width=0.9\columnwidth]{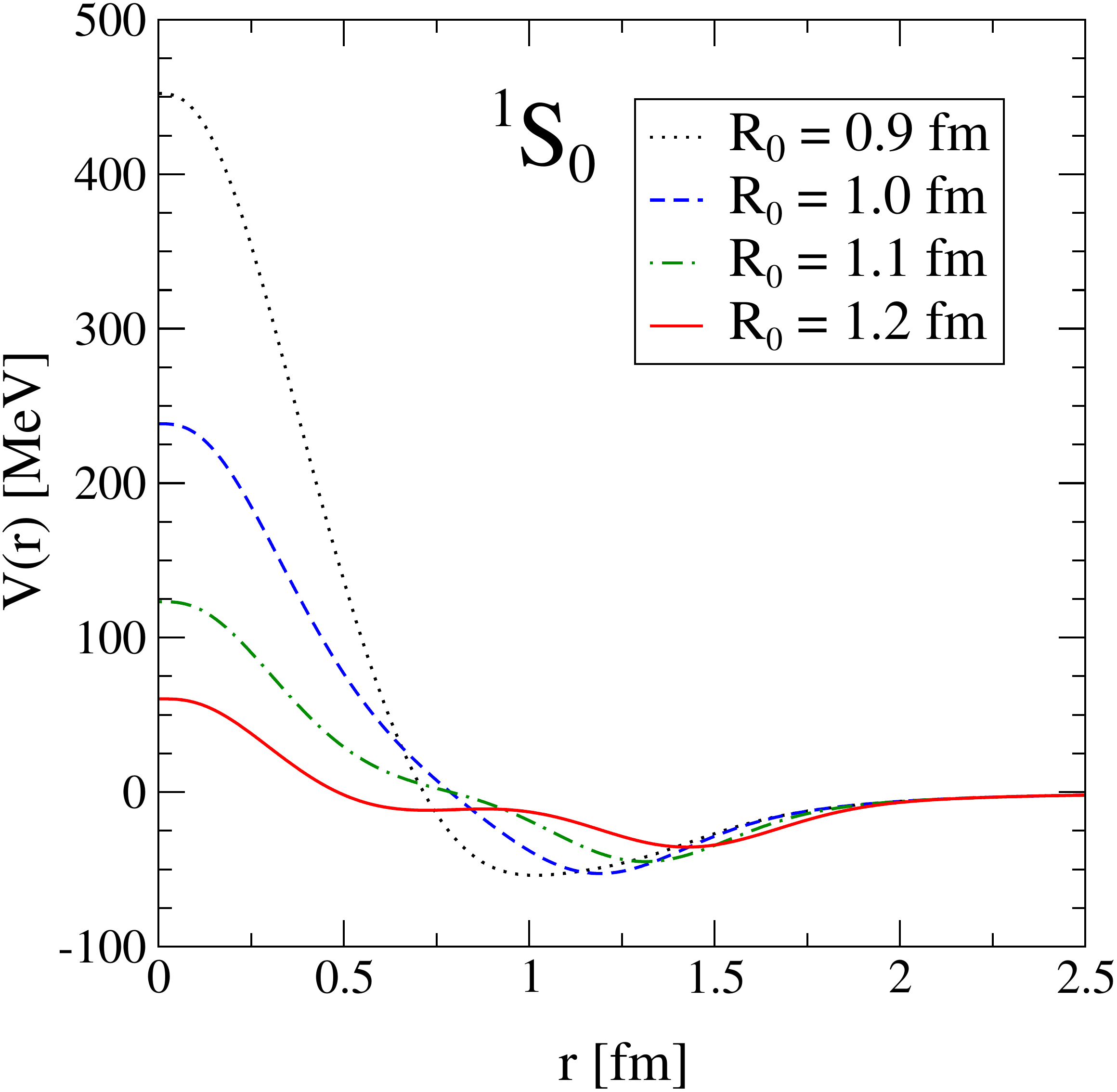}
\caption{(Color online) Local chiral NN potentials $V(r)$ at N$^2$LO
for an SFR cutoff $\tilde{\Lambda}=1000 \mev$ in the $^1S_0$ partial
wave in the neutron-neutron system.\label{fig:Pot1S0}}
\end{figure}

\begin{figure*}[t]
\centering
\includegraphics[width=0.93\textwidth]{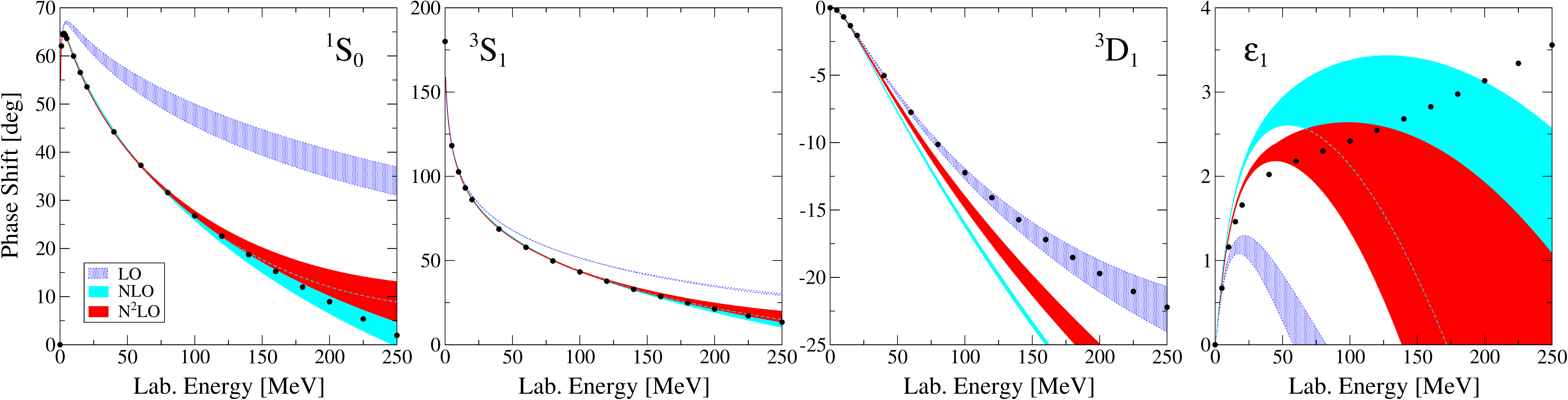}
\caption{(Color online) Phase shifts for the $^1S_0$ and $^3S_1-^3D_1$
partial waves at LO, NLO, and N$^2$LO in comparison with the Nijmegen
partial wave analysis (PWA)~\cite{Stoks:1993tb}. The bands at each
order correspond to the cutoff variation of $R_0=1.0-1.2 \fm$. At NLO
and N$^2$LO, we also vary the SFR cutoff from $\tilde{\Lambda}=1.0-1.4
\gev$.\label{fig:PS1}}
\end{figure*} 

\begin{figure*}[t]
\centering
\includegraphics[width=0.7\textwidth]{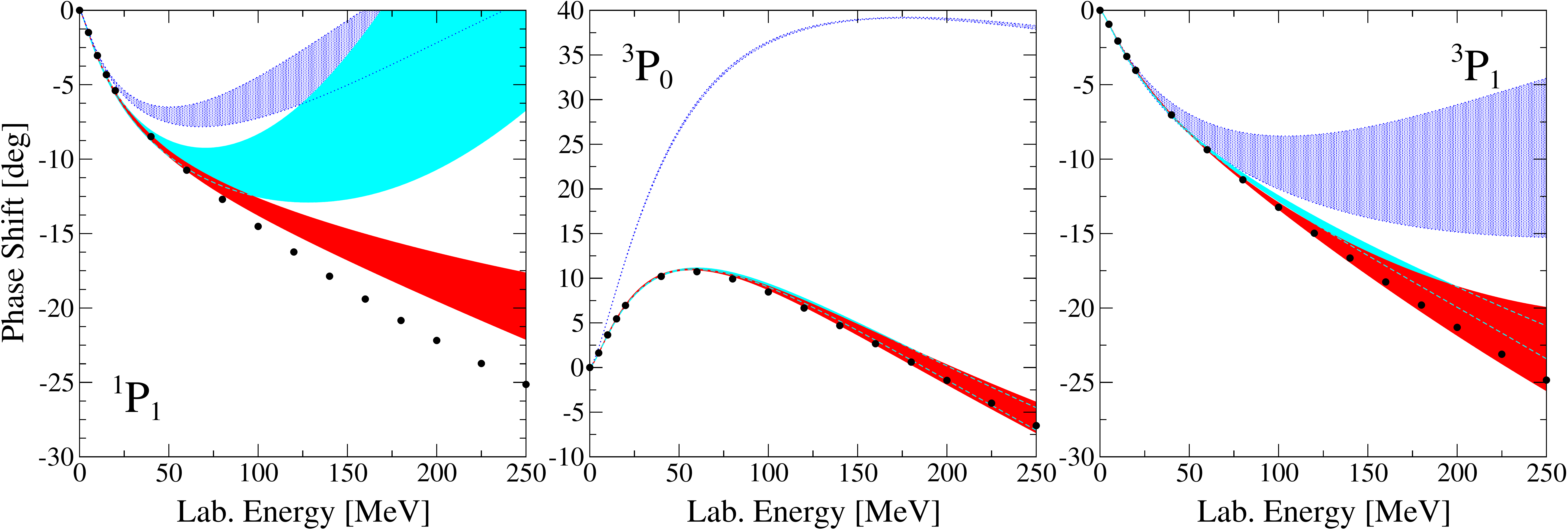}\\
\vspace{0.2cm}
\includegraphics[width=0.7\textwidth]{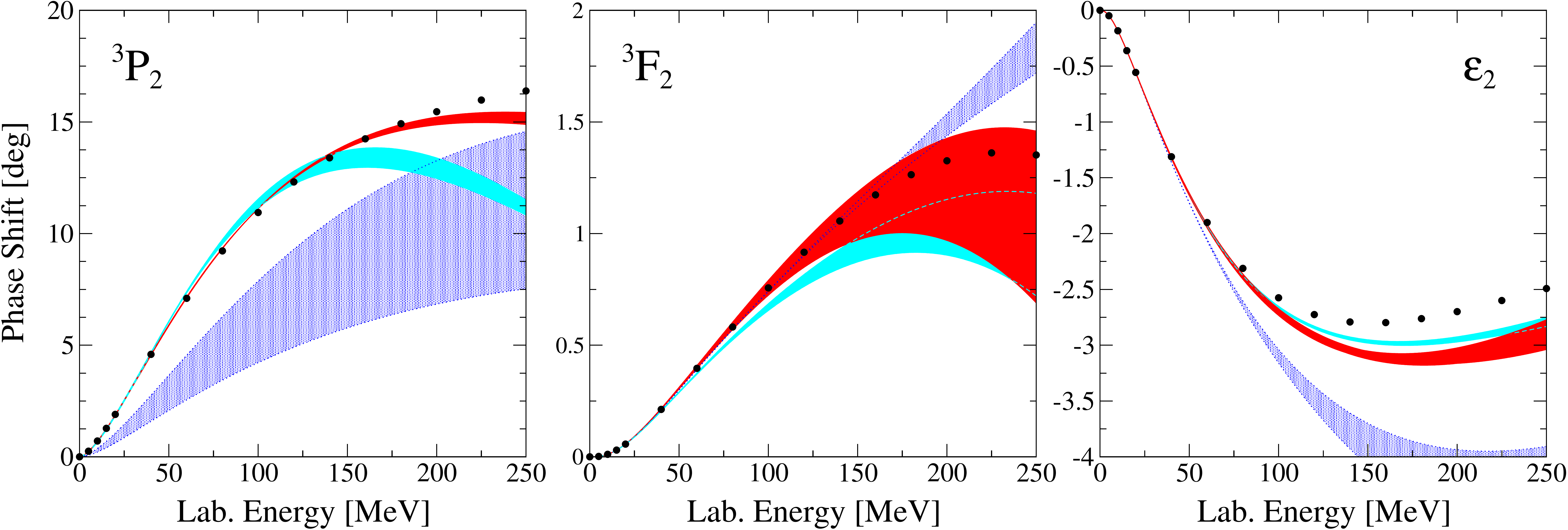}
\caption{(Color online) Phase shifts for the $^1P_1$, $^3P_0$, $^3P_1$ 
and $^3P_2$$-$$^3F_2$ partial waves at LO, NLO, and N$^2$LO in
comparison with the Nijmegen PWA~\cite{Stoks:1993tb}. The bands are
obtained as in Fig. \ref{fig:PS1}.\label{fig:PS2}}
\end{figure*}

It would be useful to have a quantitative comparison of different
fits, e.g., comparing the local chiral potentials presented here 
with the nonlocal optimized N$^2$LO potentials of
Refs.~\cite{Ekstrom:2013,Ekstrom:2014} or with the analyses of
Refs.~\cite{NavarroPerez:2014, NavarroPerez:2014b}. One possibility
would be to calculate the $\chi^2$/datum, but unfortunately we
presently do not have the machinery to do this. We also emphasize 
that our fitting strategy is different to the nonlocal optimized
N$^2$LO potentials. As discussed, we only fit at low energies and
take the $c_i$'s from pion-nucleon scattering, whereas the optimized
N$^2$LO potentials fit these over the full energy range considered.

The fits are different from the fits used in
Ref.~\cite{Gezerlis:2013ipa} because our previous fitting routine was
incorrect in the tensor channel of the pion-exchange
interactions. This error has only a small influence in pure neutron
matter.

\begin{figure*}[t]
\centering
\includegraphics[width=0.7\textwidth]{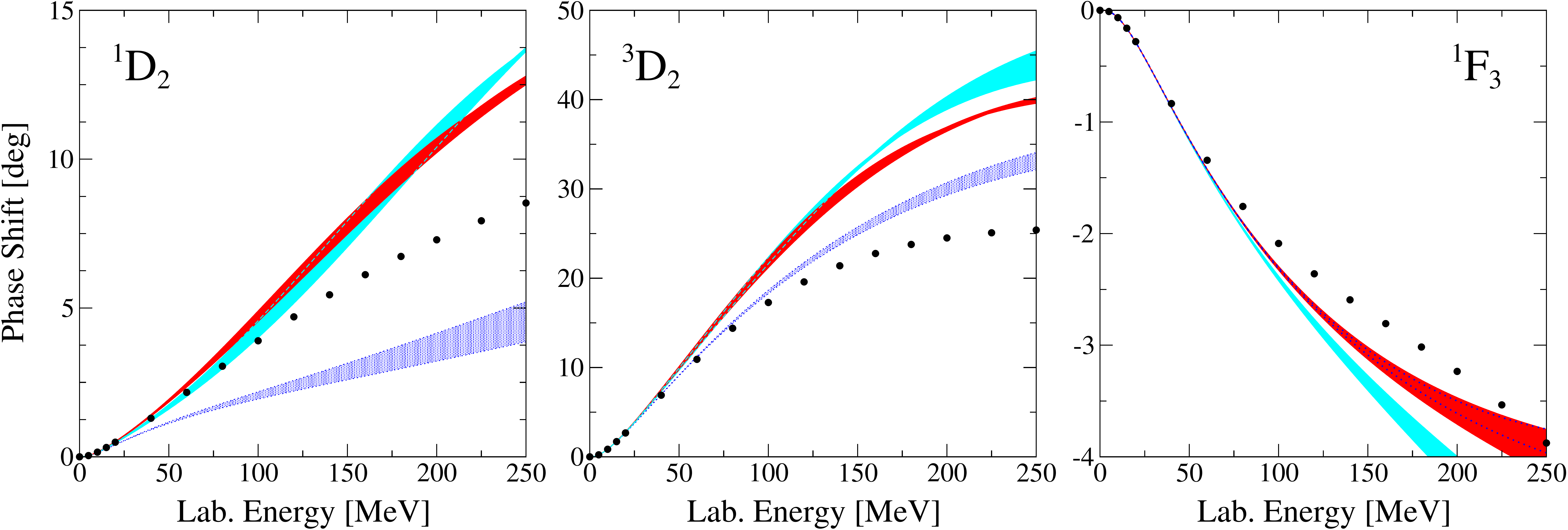}\\
\vspace{0.2cm}
\includegraphics[width=0.7\textwidth]{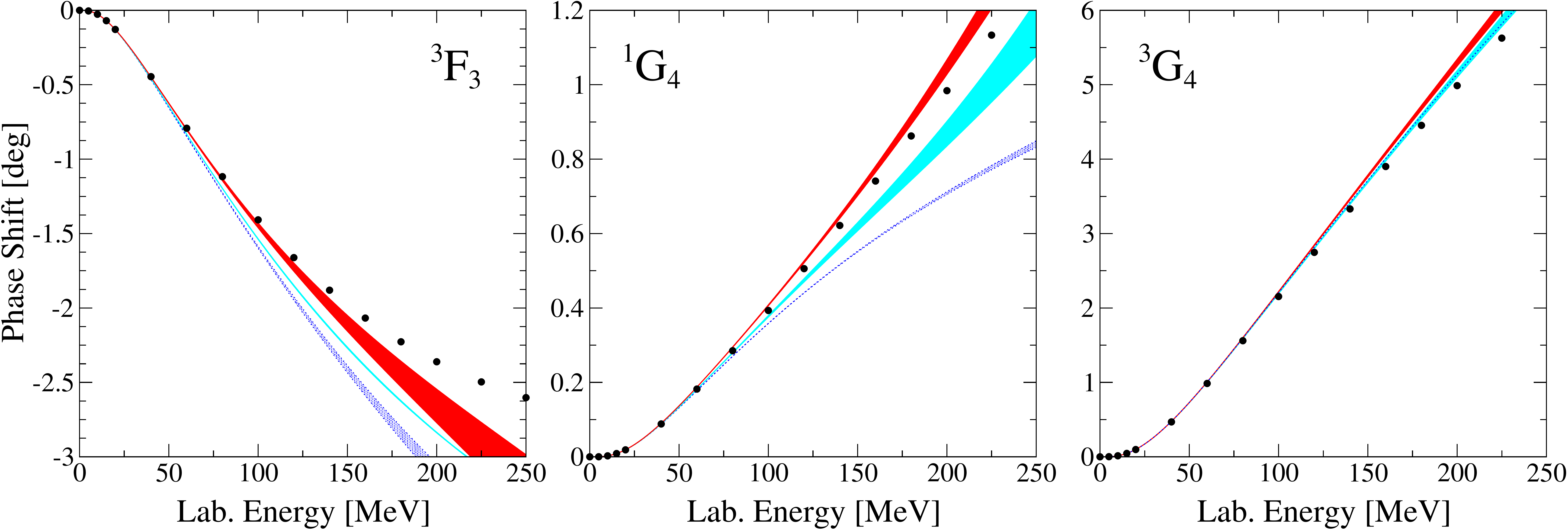}
\caption{(Color online) Phase shifts for the $^1D_2$, $^3D_2$, $^1F_3$,
$^3F_3$, $^1G_4$, and $^3G_4$ partial waves at LO, NLO, and N$^2$LO in
comparison with the Nijmegen PWA~\cite{Stoks:1993tb}. The bands are
obtained as in Fig.~\ref{fig:PS1}.\label{fig:PS3}}
\end{figure*}

\begin{figure*}[t]
\centering
\includegraphics[width=0.7\textwidth]{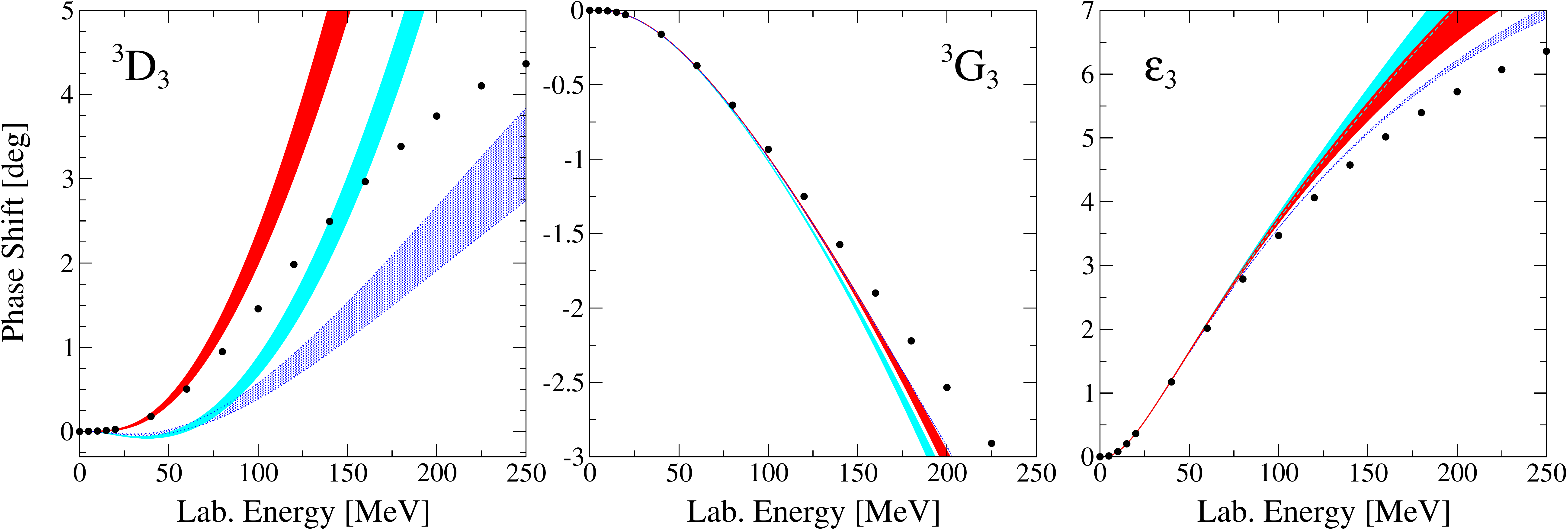}\\
\vspace{0.2cm}
\includegraphics[width=0.7\textwidth]{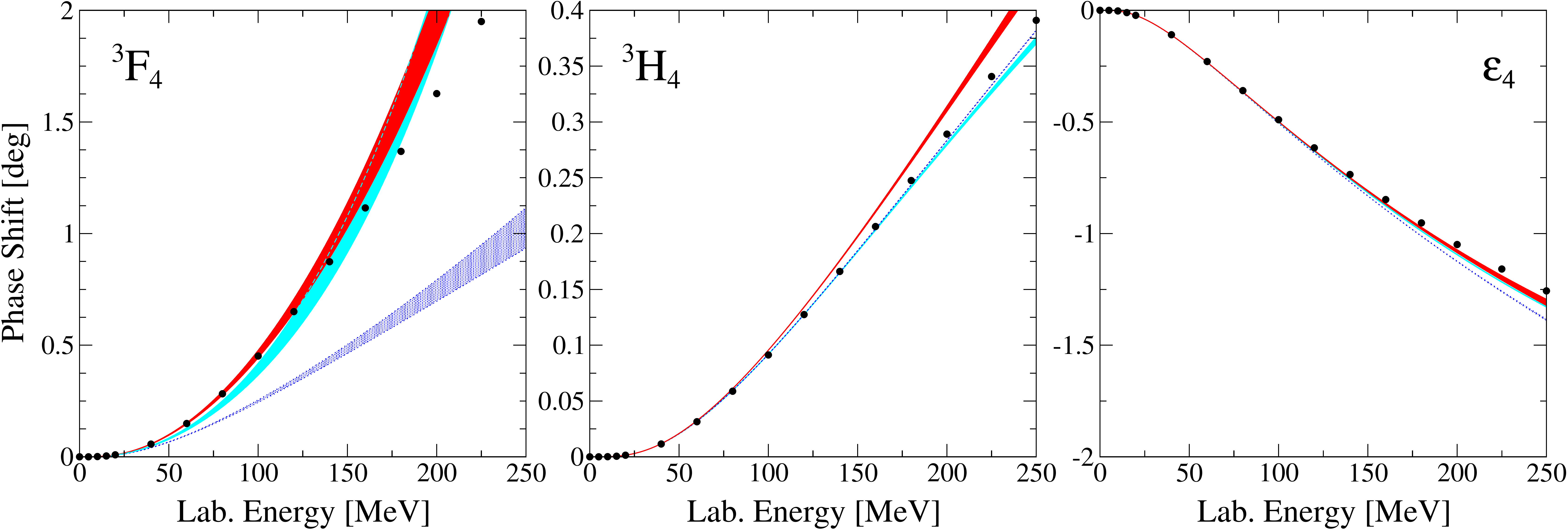}
\caption{(Color online) Phase shifts for the $^3D_3$$-$$^3G_3$ and 
$^3F_4$$-$$^3H_4$ partial waves at LO, NLO, and N$^2$LO in comparison
with the Nijmegen PWA~\cite{Stoks:1993tb}. The bands are obtained
in the same way as in Fig. \ref{fig:PS1}.\label{fig:PS4}}
\end{figure*}

In Fig.~\ref{fig:PotChannel}, we show the local chiral potentials
$V(r)$ at N$^2$LO for a SFR cutoff $\tilde{\Lambda}=1000 \mev$,
decomposed into the central, central-isospin, spin, spin-isospin,
spin-orbit, tensor, and tensor-isospin components
\begin{align}
\label{vr_decomp}
V(r)  &= V^{\text{central}}(r)+V^{\text{central-isospin}}(r)\, \, \fet \tau_1 \cdot \fet \tau_2 \nonumber \\ 
&\quad +  \left[V^{\text{spin}}(r) +  V^{\text{spin-isospin}}(r)\,\, \fet \tau_1 \cdot \fet \tau_2 \right]\fet \sigma_1 \cdot \fet \sigma_2  \nonumber \\ 
\nonumber 
&\quad +  V^{\text{LS}}(r)\,\,{\bf{L}}\cdot {\bf{S}}\nonumber \\
&\quad + \left[V^{\text{tensor}}(r)+V^{\text{tensor-isospin}}(r)\fet \tau_1 \cdot \fet \tau_2 \right]S_{12}(r) \,, 
\end{align}
for cutoffs $R_0 = 0.9-1.2 \fm$. We include the $0.9 \fm$ potential
for illustration, but we do not recommend it for many-body
calculations and therefore do not include it in our own calculations
or in the tables. For all components we see a softening of the
potential going from $R_0=0.9 \fm$ to $R_0=1.2 \fm$, as expected,
because short-range parts of the potentials are strongly scheme
dependent.  The structures in the individual channels are due to
adding up different contributions to those channels with different
$r$-dependencies.

In addition, we show the local chiral potentials $V(r)$ at N$^2$LO for
a SFR cutoff $\tilde{\Lambda}=1000 \mev$ in the $^1S_0$ channel in
Fig.~\ref{fig:Pot1S0} in the neutron-neutron system. Again, we observe
a softening of the potential when increasing the coordinate space
cutoff from $R_0=0.9 \fm$ to $R_0=1.2\fm$.

\begin{table*}[t]
\begin{center}
\caption{Deuteron properties for the local chiral potentials at 
LO, NLO, and N$^2$LO.  We tabulate the deuteron binding energy $E_d$,
the $D$-state probability $P_D$, the magnetic moment $\mu_d$, the
quadrupole moment $Q_d$, the asymptotic D/S ratio $\eta$, the
asymptotic $S$-wave factor $A_s$, and the rms radius $r_d$. The ranges
include a cutoff variation $R_0=1.0-1.2 \fm$ and, at NLO and N$^2$LO,
a variation of the SFR cutoff $\tilde{\Lambda}=1.0-1.4 \gev$. The
experimental results are taken from
Refs.~\cite{Deuteron:Eb,Deuteron:mud,Deuteron:Qd,Deuteron:eta,%
Deuteron:rd,Deuteron:AS}.  We compare our results with the N$^2$LO
EGM results of Ref.~\cite{EGMN2LO}, where the cutoff variation is
$\Lambda=450-650 \mev$ and $\tilde{\Lambda}=500-700 \mev$.\label{tab:deuteron}}
\begin{tabular}{l|c|c|c||c||c}
\hline
& LO & NLO & N$^2$LO & N$^2$LO EGM & Exp.  \\
\hline
$E_d [\mev]$ & $-2.0243 \ldots -2.0161$ & $ -2.1597 \ldots -2.1446 $ & $ -2.2177 \ldots -2.1981 $ & $-2.202 \ldots -2.189$ &$-2.225$\\
$P_D [\%]$ & $4.2761 \ldots 5.3356$ &$6.9249 \ldots 8.1702$ & $ 5.5059 \ldots 6.1356$ & $3.53\ldots4.93$ &\\
$\mu_d [\mu_N]$ & $0.8494 \ldots 0.8554$ & $0.8332 \ldots 0.8403$ & $0.8438 \ldots 0.8484$ & & $0.857$\\
$Q_d [\fm^2]$ & $0.2580 \ldots 0.2691 $ & $0.3013 \ldots 0.3039$ & $0.2828 \ldots 0.2890$ & $0.271 \ldots 0.275$ & $0.286$ \\
$\eta$ & $0.0232 \ldots 0.0240$ & $ 0.0275\ldots 0.0278 $ & $0.0256 \ldots 0.0267 $ & $0.0255\ldots 0.0256$ & $0.0256$\\
$A_s [\fm^{-\frac{1}{2}}]$ & $0.8299 \ldots 0.8321$ & $ 0.8605 \ldots 0.8648 $ & $0.8765 \ldots 0.8818$ & $0.874\ldots 0.879$ & $0.885$\\
$r_d [\fm]$ & $1.9897 \ldots 1.9919$ & $ 1.9737 \ldots 1.9758$ & $1.9677 \ldots 1.9698$ & $1.970 \ldots 1.972$ & $1.966$\\
\hline 
\end{tabular}
\end{center}
\end{table*}

\section{Phase shifts}\label{sect:phaseshifts}

Next, we present the neutron-proton phase shifts in partial waves up
to $J=4$ for the local chiral potentials at LO, NLO, and N$^2$LO for
laboratory energies up to $250 \mev$ in comparison with the Nijmegen
PWA~\cite{Stoks:1993tb}. We vary the cutoff between $R_0=1.0-1.2 \fm$
and, at NLO and N$^2$LO, the SFR cutoff between
$\tilde{\Lambda}=1.0-1.4 \gev$.

In Fig.~\ref{fig:PS1}, we show the $^1S_0$ phase shifts as well as the
$^3S_1-^3D_1$ coupled channel.  The description of the $^1S_0$ channel
at LO is only good at very low energies and improves when going to NLO
and the effective range physics is included. When going from NLO to
N$^2$LO, the cutoff bands overlap. In the $^3S_1$ channel the
situation is similar but the cutoff bands are narrower. In both
$S$-wave channels the width of the bands at NLO and N$^2$LO are of
similar size. This is due to the truncation of the short-range contact
interactions and the large $c_i$ couplings entering at N$^2$LO, and is
visible in all phase shifts.

In the $^3D_1$ channel the description worsens when going from LO to
NLO and improves only slightly from NLO to N$^2$LO. At N$^2$LO the
description of the $^3D_1$ channel is poor for energies larger than
$50 \mev$. In addition, also the description of the $J=1$ mixing angle
is poor at all orders, a fact which is clearly reflected in the size
of the cutoff bands.

In Fig. \ref{fig:PS2} we show the phase shifts for the $P$ waves and
the $J=2$ coupled channel. In the $^1P_1$ channel the LO band starts
to deviate from the data already at low energies. Including additional
spin-orbit and tensor contributions at NLO improves the description of
the $^1P_1$ channel only little.  However, the situation highly
improves when going to N$^2$LO.

In the $^3P$ waves the phase shifts improve considerably going from LO
to higher orders and the description of the $^3P$ waves at N$^2$LO is
substantially better than in our previous
fits~\cite{Gezerlis:2013ipa}. Furthermore, the description of the
$J=2$ coupled channel is considerably better than for the $J=1$
coupled channel and improves when going from LO to N$^2$LO.

In Fig.~\ref{fig:PS3} we show the phase shifts for the remaining
uncoupled partial waves up to $J=4$. The description of the individual
channels is good even at high energies except for the $D$ waves. This
can also be seen in Fig.~\ref{fig:PS4} where we show the $J=3$ and
$J=4$ coupled channels.

In general, the description of all $D$ wave channels is poor up to
N$^2$LO and does not improve when going from NLO to N$^2$LO. This is
due to the truncation of the contact interactions at N$^2$LO because
in partial waves with orbital angular momentum $L>1$ no contact
interactions contribute at this order except for regulator
effects. Thus, the $D$ wave phase shifts are described almost solely
by pion-exchange interactions and are parameter free. This can be
improved by going to N$^3$LO. The higher $L > 2$ partial waves instead
are mostly described by long-range pion-exchange interactions and
already the OPE interaction at LO describes the data well at low
energies. Thus, the higher partial waves can be well described already
at N$^2$LO.

Comparing our phase shift results to the results obtained with the
nonlocal N$^2$LO momentum space potential of Ref.~\cite{EGMN2LO}, we
find that the local potentials describe all partial waves up to $J=4$
better except for the $D$ waves. In addition, the cutoff variation is
smaller for the local chiral potentials.

\section{Deuteron}

In this section, we calculate deuteron properties using the local
chiral potentials presented in the previous sections at LO, NLO, and
N$^2$LO. We calculate the deuteron binding energy $E_d$, the
quadrupole moment $Q_d$, the magnetic moment $\mu_d$, the asymptotic
D/S ratio $\eta$, the root-mean-square (rms) radius $r_d$, the
asymptotic $S$-wave factor $A_s$, and the $D$-state probability
$P_D$. We vary the cutoff $R_0=1.0-1.2 \fm$ and, at NLO and N$^2$LO,
the SFR cutoff $\tilde{\Lambda}=1.0-1.4 \gev$. The deuteron properties
are calculated as described in Ref.~\cite{EGMN2LO}. The results are
shown in Table~\ref{tab:deuteron} and are compared with experimental
results of Refs.~\cite{Deuteron:Eb,Deuteron:mud,Deuteron:Qd,Deuteron:eta,%
Deuteron:rd,Deuteron:AS} and the N$^2$LO Epelbaum, Gl\"ockle, and Mei{\ss}ner (EGM) results of
Ref.~\cite{EGMN2LO}, where the cutoff variation is $\Lambda=450-650
\mev$ and $\tilde{\Lambda}=500-700 \mev$.

At N$^2$LO we find a deuteron binding energy of $-2.208 \pm 0.010
\mev$, which has to be compared with the experimental value of $-2.225
\mev$. Thus, the N$^2$LO result deviates from the experimental result
by less than $1\%$, which is better than $2.196 \pm 0.007$ for the
nonlocal, momentum-space N$^2$LO EGM potentials of
Ref.~\cite{EGMN2LO}. However, for those potentials the range of the
cutoff variation is different, which affects the results and
theoretical error estimates.

The description of the deuteron quadrupole moment is surprisingly good
for the local chiral potentials and the experimental result lies
within the N$^2$LO uncertainty band. Note that electromagnetic
two-body currents are not included. The results for
the N$^2$LO momentum space potentials instead deviate by $4-5
\%$. Also for the other observables the result of the local N$^2$LO
potentials deviates less than $1 \%$ from the experimental values.

\section{QMC calculations of \\neutron matter}

\begin{figure}[t]
\includegraphics[width=\columnwidth]{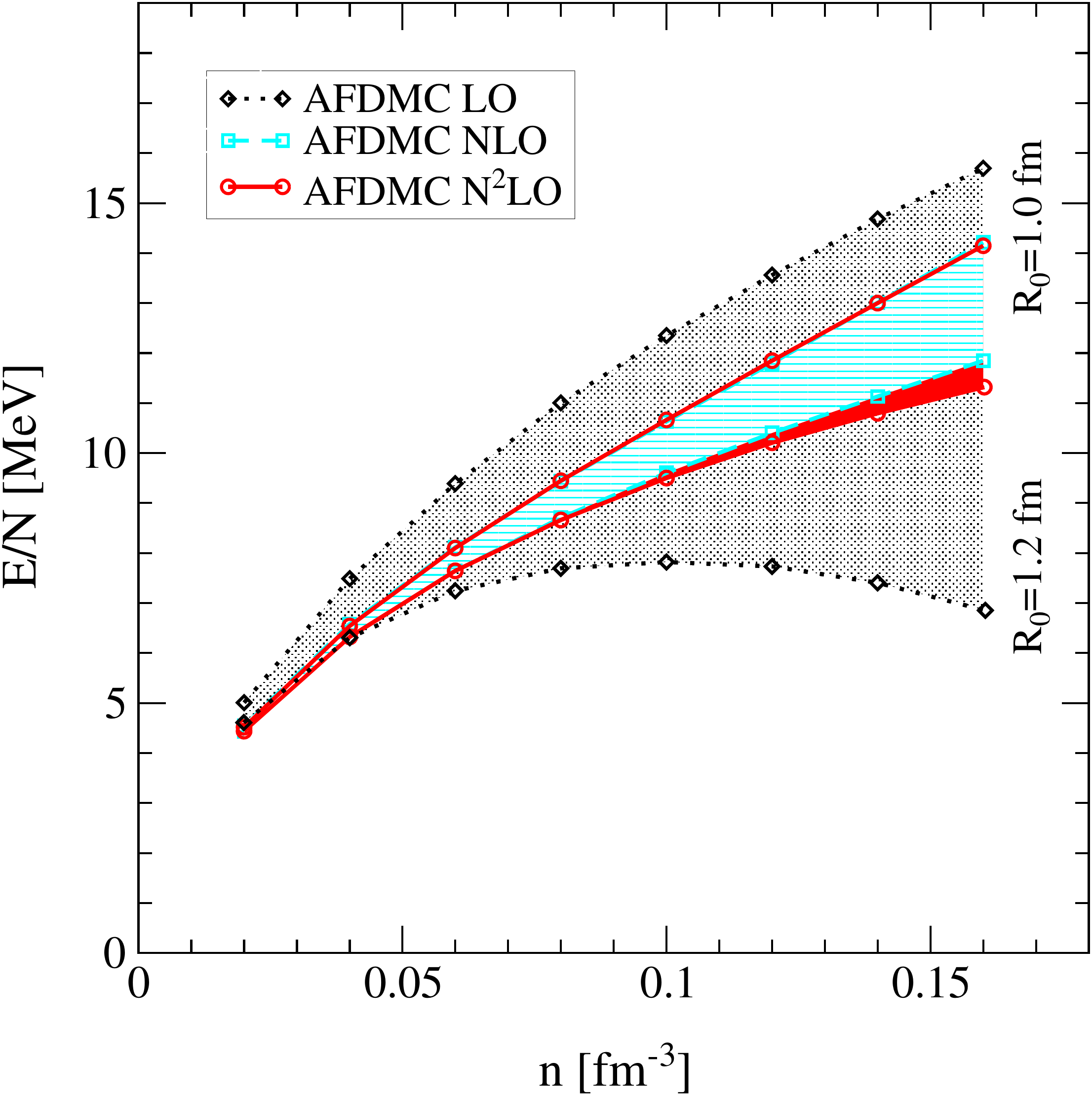}
\caption{(Color online) Neutron matter energy per particle $E/N$ 
as a function of density $n$ using AFDMC with the local chiral NN
potentials at LO, NLO, and N$^2$LO. The bands are obtained by varying
the cutoff $R_0=1.0-1.2 \fm$ and the SFR cutoff
$\tilde{\Lambda}=1000-1400 \mev$.\label{fig:AFDMCresults}}
\end{figure}

\begin{figure}[t]
\includegraphics[width=\columnwidth]{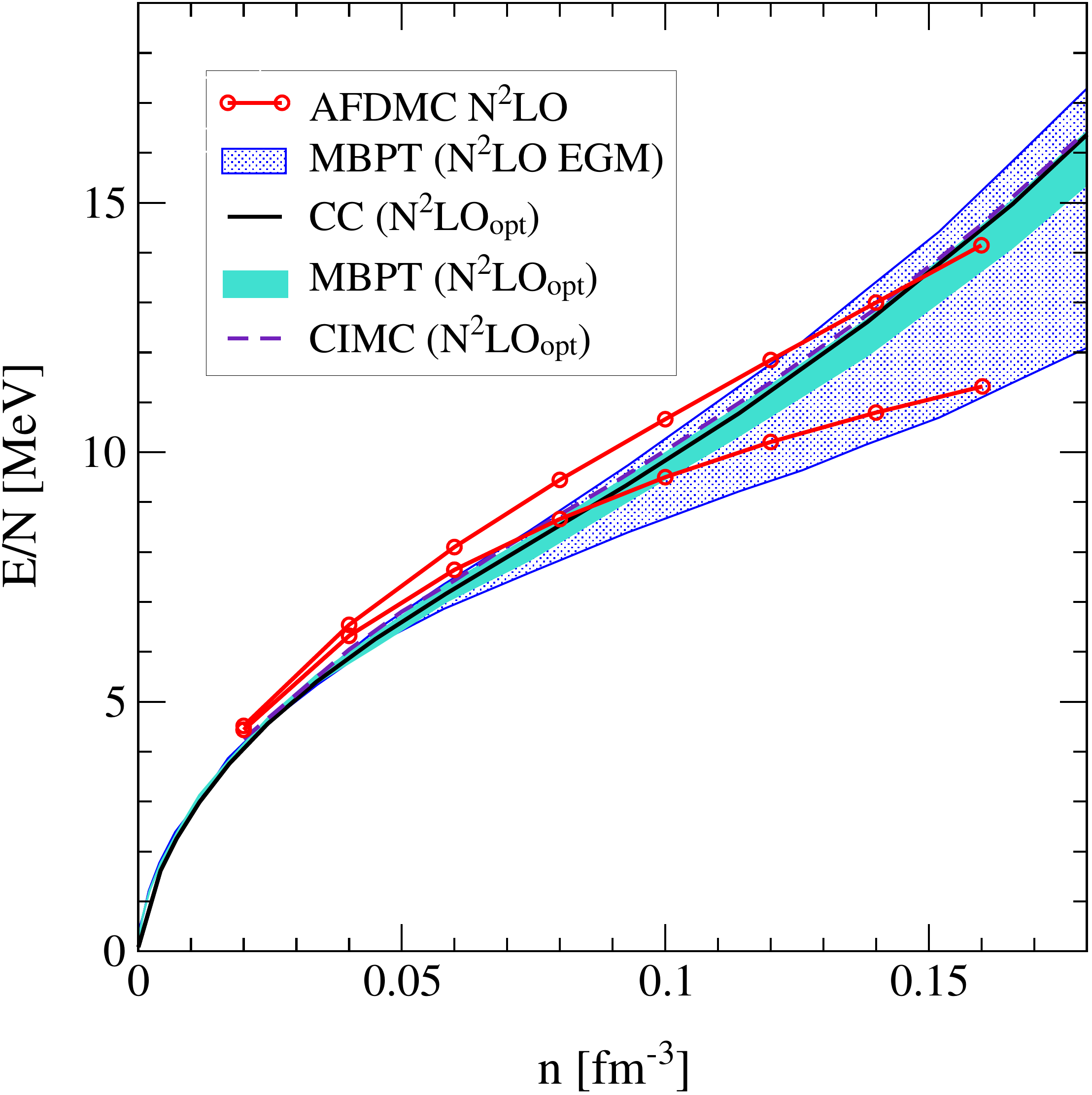}
\caption{(Color online) Neutron matter energy per 
particle $E/N$ as a function of density $n$. We compare our AFDMC
N$^2$LO results of this work with the MBPT N$^2$LO results of
Ref.~\cite{Tews:2013} using the momentum-space potentials of
Ref.~\cite{EGMN2LO}, the coupled-cluster results of
Ref.~\cite{Hagen:2014} using the optimized N$^2$LO potential of
Ref.~\cite{Ekstrom:2013}, the MBPT results of Ref.~\cite{Tews:2013b},
and the configuration interaction Monte Carlo (CIMC) results of Ref.~\cite{Roggero:2014}, both using the same
optimized N$^2$LO potential.\label{fig:AFDMCcomparison}}
\end{figure}

Local chiral EFT interactions can be used in any modern many-body
method. This includes Quantum Monte Carlo. The two main methods in the
context of nuclear physics are GFMC, which is very accurate but also
computationally costly, and AFDMC, which is computationally less
costly at the price of less accuracy. Up to now, nuclear GFMC
calculations have used phenomenological NN interactions as input,
typically of the Argonne family~\cite{Wiringa1995, Wiringa2002}. These
potentials are accurate, but are not connected to an EFT of QCD and
their two-pion exchange interaction is modeled rather
phenomenologically, which makes it difficult to construct consistent
3N forces. Thus, it will be key to use the new local potentials in
light nuclei GFMC calculations, work that is currently
ongoing~\cite{Lynn2014}.

In this paper, we use the new local chiral potentials in AFDMC
calculations for pure neutron matter and expand on our first results
of Ref.~\cite{Gezerlis:2013ipa}.  For technical reasons, in the past
it has not been possible to extend AFDMC to realistic potentials when
both neutrons and protons are involved. However, for pure neutron
matter, either in the homogeneous case or in a confining potential,
the situation is more straightforward and AFDMC compares favorably
with the more accurate nuclear GFMC results~\cite{Gandolfi:2011,%
Carlson:2003}. Neutron matter is useful as a test case in which
different aspects of nuclear interactions can be probed, but is also
directly relevant to the properties of neutron stars and as ab initio
input to energy density functionals~\cite{Gandolfi:2011,Gandolfi:2012,%
Gezerlis:2008, Chamel:2008,Hebeler:2010b,nstar_long}.

In Fig.~\ref{fig:AFDMCresults}, we show AFDMC results for 66 neutrons
for the local chiral potentials at LO, NLO, and N$^2$LO, varying $R_0
= 1.0 - 1.2 \fm$, corresponding to a cutoff range of $\sim 500 - 400
\mev$ in momentum space, and the SFR cutoff $\tilde{\Lambda} = 1000 -
1400 \mev$.  At all these orders the $R_0 = 1.1 \fm$ results lie
between the $R_0 = 1.0 \fm$ and $R_0 = 1.2 \fm$ ones. This can also be
seen in more detail in Fig.~\ref{fig:NNconvergence}, where we show the
AFDMC results individually for three different regulators $R_0
= 1.0 \fm, 1.1 \fm,$ and $1.2 \fm$ and a SFR cutoff of
$\tilde{\Lambda} = 1000 \mev$, along with the many-body perturbation
theory results that will be discussed in the next section.

As shown in Ref.~\cite{Gezerlis:2013ipa}, the LO results lead to a
broad band, the lower part of which ($R_0 = 1.2 \fm$) even changes
slope as the density is increased. This reflects the fact that the LO
potential does not describe the phase shifts at the relevant energies
as there are only two LECs at this order.  The NLO and N$^2$LO results
are generally similar in size, as observed in
Ref.~\cite{Gezerlis:2013ipa}, due to the large $c_i$ entering at
N$^2$LO and the same truncation of the contact interactions at both
orders. The width of these bands is similar to that of the phase
shifts discussed in Sect.~\ref{sect:phaseshifts}.

\begin{figure}[t]
\includegraphics[height=\columnwidth]{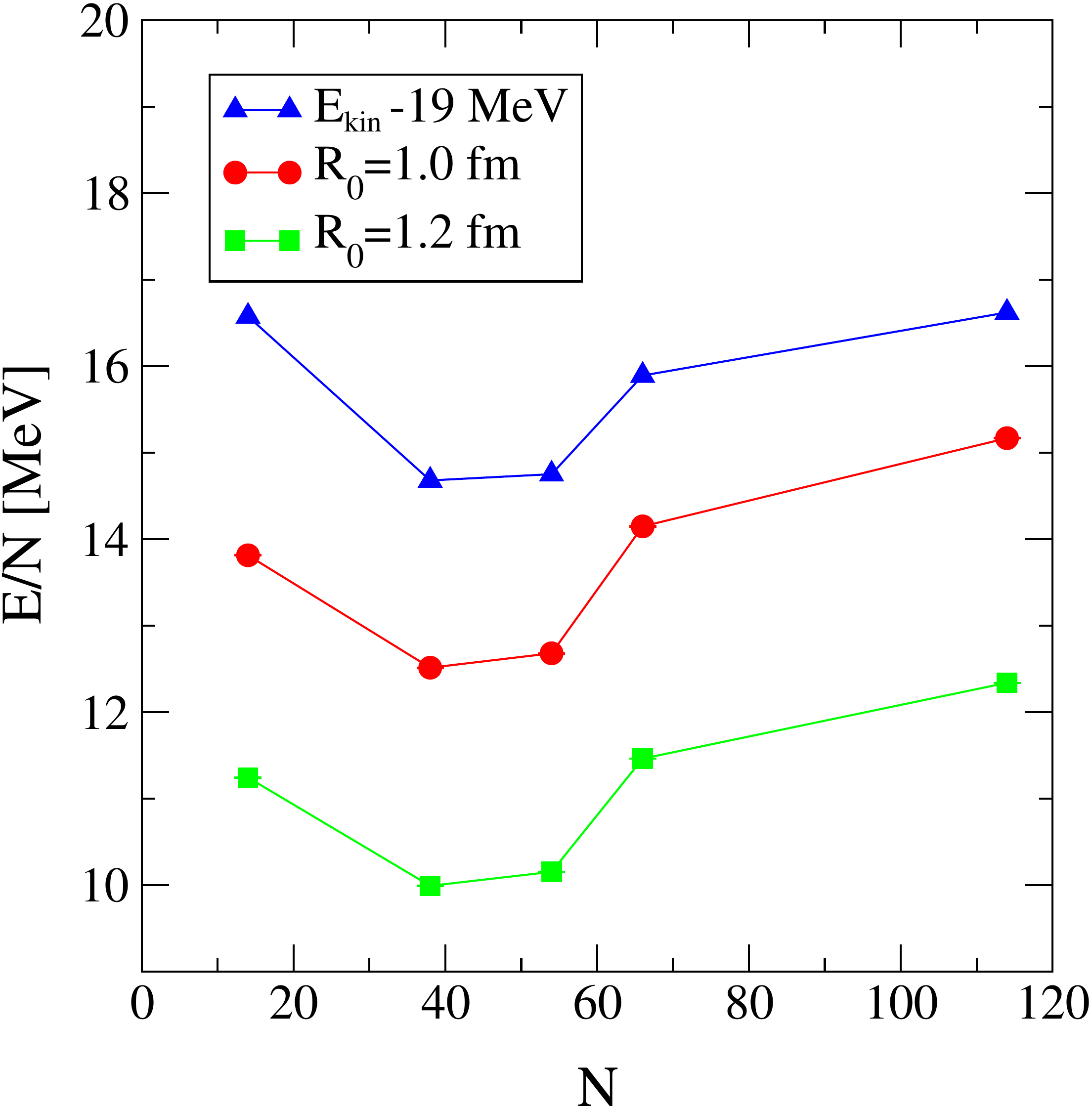}
\caption{(Color online) Finite-size effects for the ground-state 
energy of neutron matter for a SFR cutoff $\tilde{\Lambda} = 1000
\mev$ at N$^2$LO. Results are shown for different particle numbers for
the $R_0 = 1.0 \fm$ and the $R_0 = 1.2 \fm$ potentials. We also show
the kinetic energy, shifted down by $19 \mev$. The finite-size effects
for the local chiral potentials follow the shell effects of the kinetic
energy operator.\label{fig:finitesize}}
\end{figure}

\begin{figure}[t]
\includegraphics[height=\columnwidth]{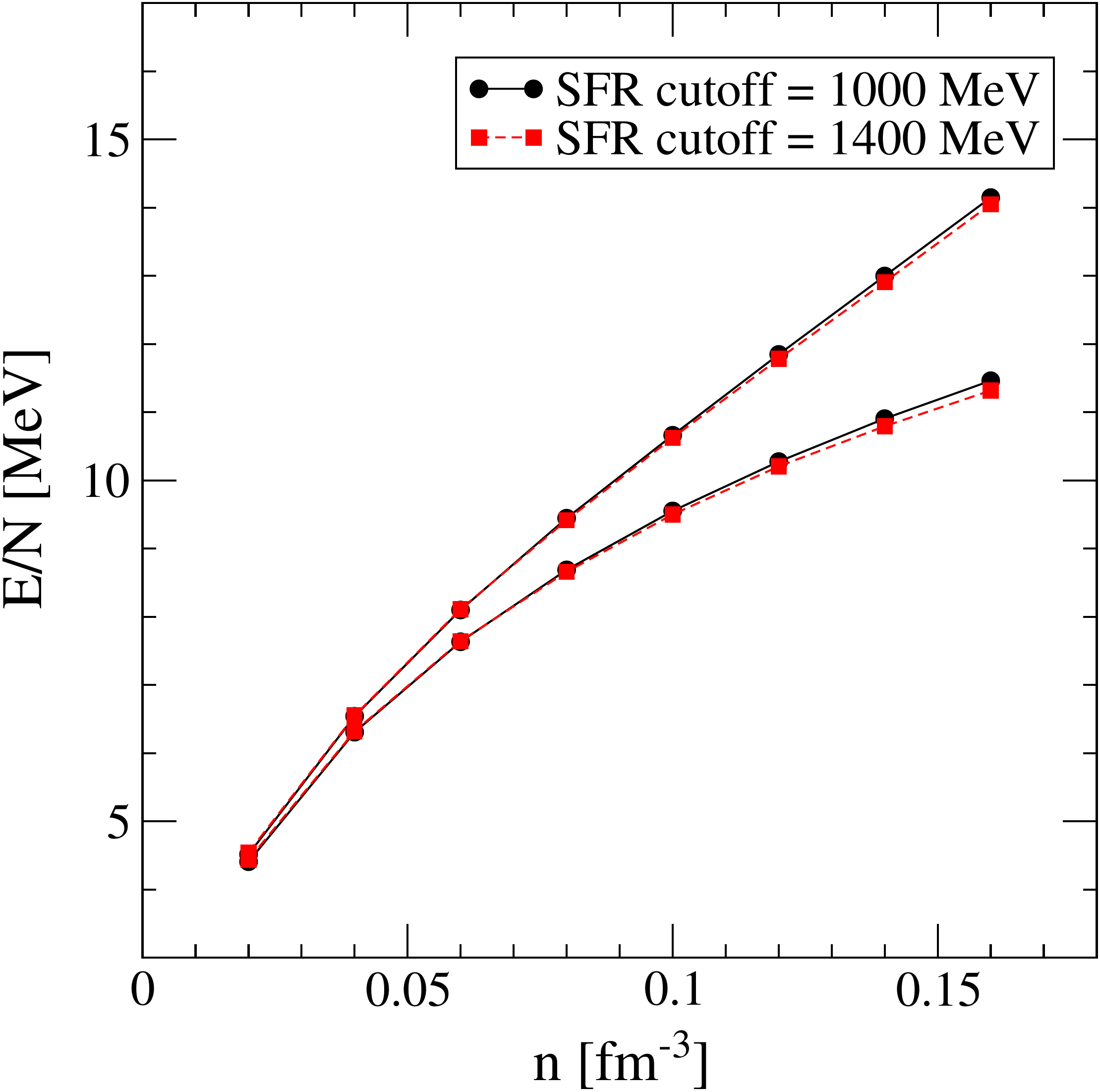}
\caption{(Color online) Ground-state energy of 66 neutrons at N$^2$LO.
Shown are results for two SFR cutoffs, $\tilde{\Lambda} = 1000 \mev$ 
and $\tilde{\Lambda} = 1400 \mev$, and two different cutoffs $R_0 =
1.0 \fm$ (upper lines) and $R_0 = 1.2 \fm$ (lower lines). The results
exhibit a very weak $\tilde{\Lambda}$ dependence.\label{fig:SFR}}
\end{figure}

In Ref.~\cite{Gezerlis:2013ipa}, we varied the cutoff from $R_0 = 0.8
\fm$ to $R_0 = 1.2 \fm$. Since we have been unable to produce a
precision potential with no deeply bound states for $R_0 = 0.8 \fm$,
we cannot directly compare our new AFDMC results with those of
Ref.~\cite{Gezerlis:2013ipa}, because the latter had an error in 
the fitting routine for the tensor channel of the pion-exchange 
interactions, which however only has a small effect on pure neutron 
matter. The narrower range of cutoff variation in this work has made 
the bands somewhat smaller, at $0.15 \fm^{-3}$, the range is
 $8.1 \mev$ at LO, $2.1 \mev$ at NLO, and $2.1 \mev$ at N$^2$LO.
 
In Fig.~\ref{fig:AFDMCcomparison} we compare our 
AFDMC N$^2$LO results for neutron matter with the MBPT N$^2$LO calculation 
of Ref.~\cite{Tews:2013} based on the momentum-space potentials of
Ref.~\cite{EGMN2LO}, the coupled-cluster results of Ref.~\cite{Hagen:2014} 
using the optimized N$^2$LO potential of Ref.~\cite{Ekstrom:2013}, 
the MBPT results of Ref.~\cite{Tews:2013b}, and the CIMC calculation of
Ref.~\cite{Roggero:2014}, both using the same optimized N$^2$LO potential.
The bands for the MBPT results are obtained as described in 
Ref.~\cite{Tews:2013}.

The different many-body results for the optimized N$^2$LO potential are 
in very good agreement. These results also
agree very well with recent self-consistent Green's function
results~\cite{Carbone:2014}. In addition, the optimized N$^2$LO
results agree very well with the N$^2$LO band of Ref.~\cite{Tews:2013}
which includes also a NN cutoff variation and is therefore rather
broad. Comparing with the AFDMC results of this paper, we find that at
saturation density the resulting energies per particle agree very
well. However, the general density dependence of the AFDMC results is
more flat, leading to higher energies at intermediate densities and a
different density dependence at saturation density. These differences 
could be due to the differences in the phase shift predictions, 
and we expect both results to come closer when going to N$^3$LO.
 
We have also tested the dependence of the AFDMC results on the Jastrow
term in the variational wave function. Specifically, the trial wave
function in AFDMC is written as 
\begin{equation}
\Psi_T(\mathbf{R},\mathbf{S}) = {\cal A} \left[\prod_{i} \phi_{\alpha}(\mathbf{r}_i,s_i)\right] \prod_{i<j} f(r_{ij}) \,,
\label{eq:AFDMCwave}
\end{equation}
where $\alpha$ labels the single-particle state.
For a nodeless Jastrow term, most QMC methods are independent of the
choice one makes for $f(r)$: the Jastrow function impacts the
statistical error bar by accenting the ``appropriate'' regions of
phase space, but not the value itself. However, due to the complicated
spin-dependence of nuclear interactions, it has been found that AFDMC
has a small dependence on the Jastrow function as reported in
Ref.~\cite{Gezerlis:2013ipa}. By comparing AFDMC results for 14
particles using the Argonne family of potentials with a GFMC
calculation for the same potentials and neutron number (the largest
neutron number for which GFMC results exist), we found that the
Jastrow dependence disappears in AFDMC when using a softened Jastrow
function.

Because no GFMC results exist for 66 particles, we have carried out
separate computations at the highest density considered here ($n =
0.16 \fm^{-3}$). We studied Jastrow terms from solving the
Schr\"odinger equation for the Argonne $v_8'$ potential, a typical QMC
potential of reference, and from the consistent local chiral
potentials. In addition, we have examined the effect of artificially
softening the Jastrow term by multiplying the input potential (only
when producing the Jastrow function) by a fixed coefficient, in order
to see the effect of removing the Jastrow. The highest energies always
result from using a largely unmodified Argonne $v_8'$ potential, as
this is the potential that is most different from the new chiral
interactions.  In the case of $R_0 = 1.0 \fm$ the different Jastrow
terms lead to an energy per particle that varies by at most 0.1 MeV at
$0.16 \fm^{-3}$, while for the $R_0 = 1.2 \fm$ potentials the
variation is 0.15 MeV. Both these results are much smaller than the
$0.6 \mev$ quoted in Ref.~\cite{Gezerlis:2013ipa} for the $R_0 = 0.8
\fm$ potential. This is a reflection of the softer potentials in the
present work.

Furthermore, we have probed in detail the finite-size effects for the
local chiral potentials. As we are interested in describing the
thermodynamic limit of neutron matter, it is important that we are
using sufficiently many particles in our AFDMC simulations.  In order
to avoid issues related to preferred directions in momentum-space, we
have performed calculations for closed shells: $N= 14$, 38, 54, 66,
114. We chose the SFR cutoff $\tilde{\Lambda} = 1000 \mev$ and
performed simulations at N$^2$LO for both the $R_0 = 1.0 \fm$ and $R_0
= 1.2 \fm$ potentials at the highest density $n=0.16 \fm^{-3}$. The
results are shown in Fig.~\ref{fig:finitesize}. We observe that the
two potentials exhibit essentially identical shell structure, as was
to be expected because the ranges involved in the two potentials are
basically the same. These results show a dependence on $N$ that is
very similar to that in Table III of Ref.~\cite{Gandolfi:2009} for the
values of $N$ used in that reference, namely 14, 38, and 66. The shell
structure is very similar to that of the free Fermi gas in a periodic
box, which we also show in Fig.~\ref{fig:finitesize}. From the free
Fermi gas we expect that the thermodynamic limit value is below the
$N=114$ result and very close to the $N=66$ value. This justifies our
choice of using 66 particles to simulate the thermodynamic limit. The
only qualitative difference between the free Fermi-gas shell structure
and our AFDMC results appears at $N=14$. For the free gas $N=14$ leads
to an energy that is higher than that at $N=66$. This results from the
very small periodic box needed to produce the same density for
$N=14$. In that case the interaction length scales also start to be
important. In contrast, for larger $N$, shell effects come almost
completely from the kinetic energy behavior.

We have also explored the dependence of the results on different
values of the SFR cutoff. As discussed, the effect of the SFR cutoff
$\tilde{\Lambda}$ is expected to be smaller than that of $R_0$. We
show the results of varying the SFR cutoff from $\tilde{\Lambda} =
1000$ MeV to $\tilde{\Lambda} = 1400$ MeV for $R_0 = 1.0 \fm$ and $R_0
= 1.2 \fm$ in Fig.~\ref{fig:SFR}. There is essentially no effect at
low densities, while at higher densities the difference for $R_0 = 1.0
\fm$ never exceeds 0.1 MeV and for $R_0 = 1.2 \fm$ it is always less
than 0.15 MeV.  This shows that the SFR cutoff has a negligible impact
on the many-body results.
  
\section{Perturbative calculations of \\neutron matter}

\begin{figure*}[p]
\centering
\includegraphics[width=0.99\textwidth]{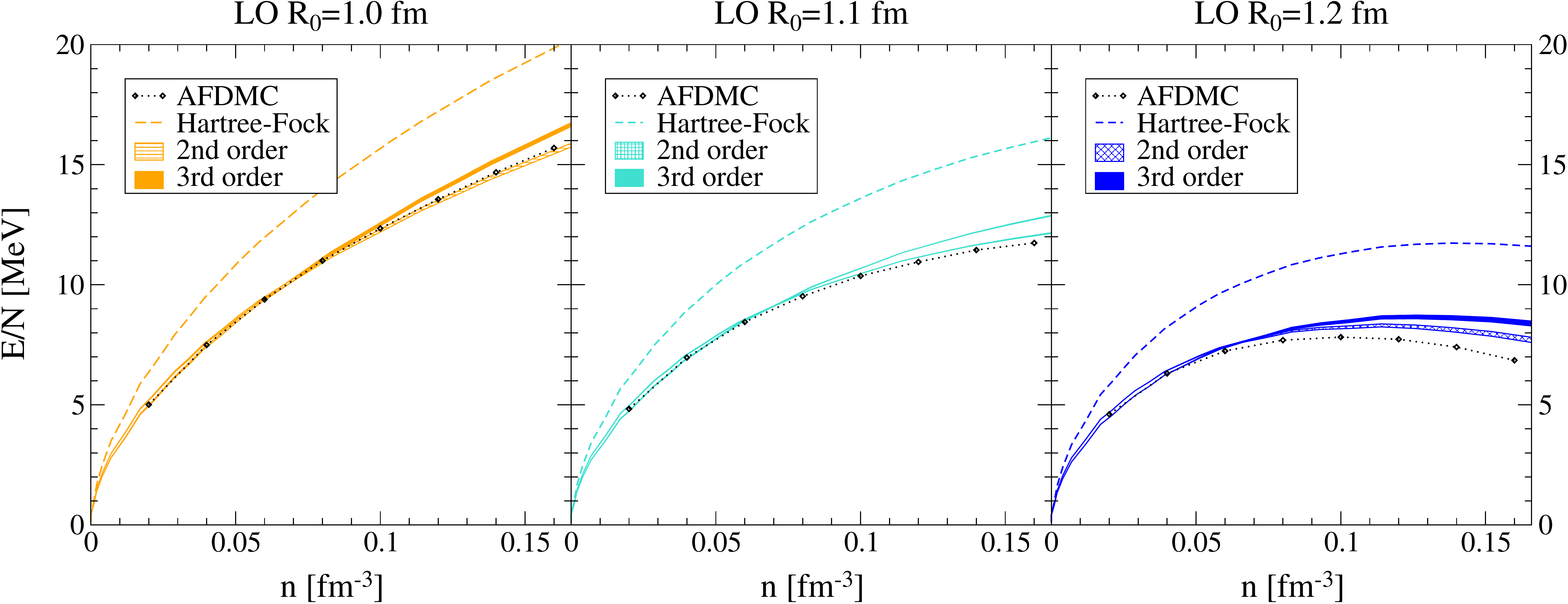}\\
\vspace{0.2cm}
\includegraphics[width=0.99\textwidth]{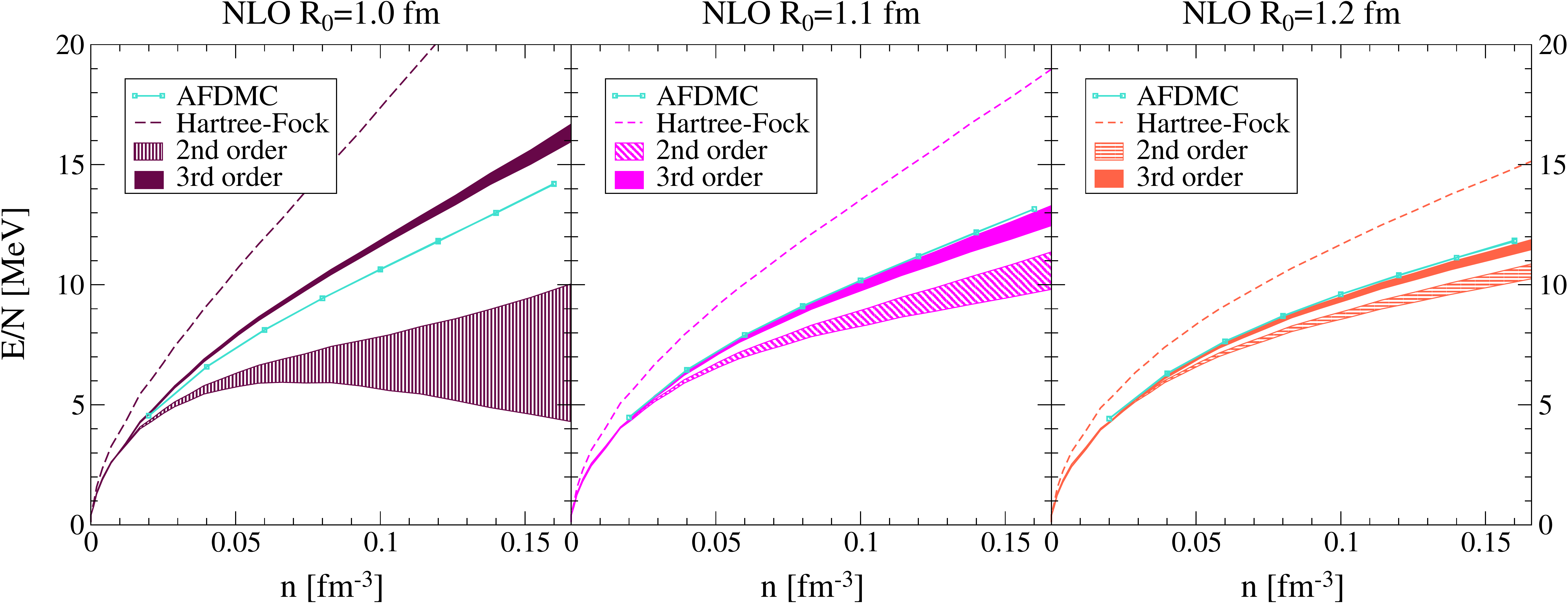}\\
\vspace{0.2cm}
\includegraphics[width=0.99\textwidth]{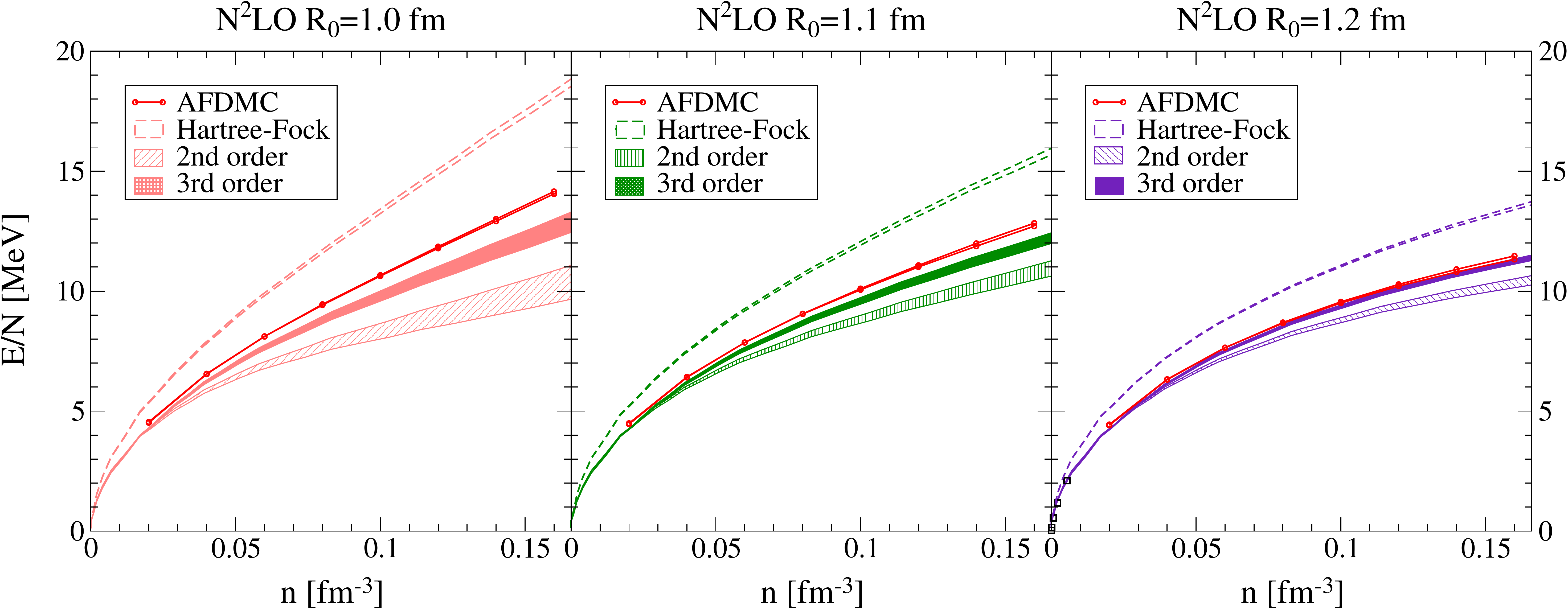}
\caption{(Color online) Results for MBPT and AFDMC calculations at 
LO, NLO, and N$^2$LO for $R_0 = 1.0 -1.2 \fm$. For the MBPT results,
we show the Hartree-Fock energies as well as the energy at second
order and including third-order particle-particle and hole-hole
corrections. The width of the bands includes a variation of the
single-particle spectrum from a free to a Hartree-Fock spectrum. In
addition, for both the MBPT and AFDMC results we also vary the SFR
cutoff $\tilde{\Lambda}=1000-1400 \mev$. For the LO $1.1 \fm$ results,
the lower band corresponds to the second-order results.
\label{fig:NNconvergence}}
\end{figure*}

We have also performed neutron matter calculations using many-body
perturbation theory (following Refs.~\cite{Hebeler:2010a,%
nucmatt,Tews:2013, Kruger:2013}) for the same local chiral
potentials and the same regulators as in the previous section. We show
the results in Fig.~\ref{fig:NNconvergence} together with the AFDMC
results at LO, NLO, and N$^2$LO for the three different cutoffs $R_0 =
1.0, 1.1,$ and $1.2 \fm$, and varying the SFR cutoff
$\tilde{\Lambda} = 1000-1400 \mev$.

At every order in the chiral expansion and for every cutoff we show
the results at the Hartree-Fock level as a dashed line, including
second-order contributions as a shaded band, and including also
third-order particle-particle and hole-hole corrections as solid
bands. The bands are obtained by employing a free or Hartree-Fock
single-particle spectrum and by varying the SFR cutoff as stated
above. Again, we observe that the $R_0 = 1.1 \fm$ results at all three
chiral orders lie between the $R_0 = 1.0 \fm$ and $R_0 = 1.2 \fm$
ones.

At LO, the local chiral potentials in general follow the trend of the
AFDMC results for all three cutoffs. The width of the individual bands
is very small and the energy changes from first to second and from
second to third order are small. As discussed in
Ref.~\cite{Kruger:2013}, this energy difference, combined with the
weak dependence on the different single-particle spectra, is a measure
of the perturbative convergence for the individual potentials. All
potentials at this chiral order seem to be perturbative. We find a
good agreement between the AFDMC and the MBPT results, especially at
lower densities, although at higher densities the trend is that the
second-order results are better than third-order.

At NLO, we find the $R_0 = 1.0 \fm$ potential to have the slowest, if
any, perturbative convergence. The second-order band is very broad and
the third-order contributions are large: at saturation density they
are $6-10 \mev$. Going to higher coordinate-space cutoffs, which
means lower momentum cutoffs, we find that the potential becomes more
perturbative. At $R_0 = 1.2 \fm$ both the second- and third-order
bands are narrow and the third-order contributions are $\approx 1.5 \mev$.

At N$^2$LO the results are very similar to NLO. We find that the $R_0
=1.0 \fm$ potential shows the slowest perturbative convergence, with
an energy difference from second to third order of about $3 \mev$ at
saturation density. However, the perturbativeness for this cutoff at
N$^2$LO is better than at NLO. Going to higher coordinate-space
cutoffs again improves the perturbativeness and for $R_0=1.2 \fm$ the
energy difference is $\approx 1.0 \mev$ at this density. This behavior
is similar to the nonlocal potentials used in Ref.~\cite{Kruger:2013}
where it was shown that soft (low momentum cutoff) potentials have a
better convergence.

For the perturbative $R_0=1.2 \fm$ potentials, the agreement between
the third-order perturbative results and the AFDMC results is
excellent.  For $R_0 =1.2\fm$, at N$^2$LO, the perturbative results
lie almost on top of the AFDMC values. The difference between the
third-order result with Hartree-Fock single-particle spectrum and the
AFDMC results is $0.2 \mev$ at $0.16 \fm^{-3}$ for $\tilde{\Lambda} =
1400 \mev$ and only $20 \kev$ for $\tilde{\Lambda} = 1000 \mev$. In
comparison, at NLO the difference is $0.2 \mev$ at $0.16 \fm^{-3}$ for
$\tilde{\Lambda}=1400 \mev$ and $0.1 \mev$ for $\tilde{\Lambda} =1000
\mev$, while at LO it is $1.6 \mev$. These results constitute a direct
validation of MBPT for neutron matter based on low momentum
potentials, in this case $R_0 = 1.1 \fm$ and $R_0 = 1.2 \fm$, which
was the main finding in our initial QMC study with chiral EFT
interactions~\cite{Gezerlis:2013ipa}.

\section{Summary and Outlook}

We have presented details of the derivation of local chiral EFT
potentials at LO, NLO, and N$^2$LO. We performed fits of the LECs to
low-energy NN phase shifts, which are well reproduced in most cases,
and agree better than for the momentum-space potentials with the
Nijmegen PWA. Furthermore, the calculated deuteron properties at
N$^2$LO show very good agreement with experimental data.

We have applied the new local chiral potentials to neutron matter
using AFDMC and MBPT. In particular, we have investigated the
sensitivity of the results to the local regulator and to the SFR
cutoff, to the influence of the Jastrow term, and also to finite size
effects in AFDMC.

The excellent agreement of the results for the softer $R_0 = 1.1 \fm$
and $R_0 = 1.2 \fm$ potentials within the two many-body frameworks
represents a direct validation of MBPT for neutron matter and will
enable novel many-body calculations of nuclei and matter within QMC
based on chiral EFT interactions.

\begin{acknowledgments}
We thank N.~Barnea, J.~Carlson, T.~Kr\"uger, J.~Lynn, and
K.~Schmidt for useful discussions. This work was supported in
part by the European Research Council (ERC)
 Grant No.~307986 STRONGINT, by the Helmholtz Alliance
Program of the Helmholtz Association Contract No.~HA216/EMMI ``Extremes of
Density and Temperature: Cosmic Matter in the Laboratory'', by the Natural Sciences and Engineering Research Council of Canada, ERC Grant No.~259218 NuclearEFT, the US DOE SciDAC-3 NUCLEI project,
the Los Alamos National Laboratory LDRD program, and the EU HadronPhysics3
project ``Study of strongly interacting matter.'' Computations were performed at the
J\"{u}lich Supercomputing Center and at NERSC.
\end{acknowledgments}

\newpage
\appendix
\section{\\Partial-wave-decomposed contact interactions}
\label{PWD}

We fit the LECs $C_S, C_T,$ and $C_{1-7}$ to NN phase shifts. In every partial 
wave only certain LECs contribute. In the following we give the partial 
wave decomposition for all relevant channels. We use spectroscopic LECs given in terms of  
$C_S, C_T,$ and $C_{1-7}$ as follows:
\begin{align}
d_{11}&=C_S+C_T \,, \nonumber \\
d_{22}&=C_S-3C_T \,, \nonumber \\
d_{1}&=C_1-3C_2+C_3-3C_4 \,,\nonumber \\
d_{2}&=C_6-3C_7 \,,\nonumber \\
d_{3}&=C_1+C_2-3C_3-3C_4 \,,\nonumber \\
d_{4}&=C_1+C_2+C_3+C_4 \,,\nonumber \\
d_{5}&=C_1-3C_2-3C_3+9C_4 \,,\nonumber \\
d_{6}&=\frac12 C_5 \,,\nonumber \\
d_{7}&= C_6+C_7\,.\nonumber 
\end{align}

For the partial-wave-decomposed matrix elements we find
\begin{align}
\bra{^1S_0}V_{\text{cont}}\ket{^1S_0}&= d_{22}\delta_{R_0}+(d_3-d_7)\, 20 \frac{r^2}{R_0^4}\delta_{R_0}\nonumber \\& \quad-(d_3-d_7) \, 16 \frac{r^6}{R_0^8}\delta_{R_0} \,,\\
\bra{^3S_1}V_{\text{cont}}\ket{^3S_1}&= d_{11}\delta_{R_0}+(d_1+\frac13 d_2)\, 20 \frac{r^2}{R_0^4}\delta_{R_0}\nonumber \\& \quad-(d_1+\frac13 d_2)  \, 16 \frac{r^6}{R_0^8}\delta_{R_0}\,,\\
\bra{^3S_1}V_{\text{cont}}\ket{^3D_1}&= \bra{^3D_1}V_{\text{cont}}\ket{^3S_1} \\&= d_2 \frac{\sqrt{8}}{3}\, 8\frac{r^2}{R_0^4}\delta_{R_0}-d_2 \frac{\sqrt{8}}{3}\, 16\frac{r^6}{R_0^8}\delta_{R_0}\,,\nonumber \\
\bra{^3D_1}V_{\text{cont}}\ket{^3D_1}&= d_{11}\delta_{R_0}-(d_1-\frac13 d_2)  \, 16 \frac{r^6}{R_0^8}\delta_{R_0} \nonumber\\& \quad + (d_1+\frac35 d_6 +\frac{1}{15}d_2)\, 20 \frac{r^2}{R_0^4}\delta_{R_0}\,,\\
\bra{^1P_1}V_{\text{cont}}\ket{^1P_1}&= \,d_{22}\delta_{R_0}+(d_5- d_2)\, 20 \frac{r^2}{R_0^4}\delta_{R_0}\nonumber \\& \quad -(d_5- d_2)  \, 16 \frac{r^6}{R_0^8}\delta_{R_0}\,,\\
\bra{^3P_0}V_{\text{cont}}\ket{^3P_0}&= d_{11}\delta_{R_0}-(d_4-d_7)  \, 16 \frac{r^6}{R_0^8}\delta_{R_0} \nonumber \\& \quad + (d_4+\frac25 d_6 -\frac{1}{5}d_7)\, 20 \frac{r^2}{R_0^4}\delta_{R_0}\,,\\
\bra{^3P_1}V_{\text{cont}}\ket{^3P_1}&= d_{11}\delta_{R_0}-(d_4+d_7)  \, 16 \frac{r^6}{R_0^8}\delta_{R_0} \nonumber \\& \quad + (d_4+\frac15 d_6 +\frac35 d_7)\, 20 \frac{r^2}{R_0^4}\delta_{R_0}\,, \\
\bra{^3P_2}V_{\text{cont}}\ket{^3P_2}&= d_{11}\delta_{R_0}-(d_4+\frac15 d_7)  \, 16 \frac{r^6}{R_0^8}\delta_{R_0} \nonumber\\& \quad + (d_4-\frac15 d_6 +\frac{7}{25} d_7)\, 20 \frac{r^2}{R_0^4}\delta_{R_0}\,,\\
\bra{^3P_2}V_{\text{cont}}\ket{^3F_2}&= \bra{^3F_2}V_{\text{cont}}\ket{^3P_2} \\&= \, d_7 \sqrt{6}\frac{16}{5}\frac{r^2}{R_0^4}\delta_{R_0}-d_7 \sqrt{6}\frac{32}{5} \frac{r^6}{R_0^8}\delta_{R_0}\nonumber\,,\\
\bra{^3F_2}V_{\text{cont}}\ket{^3F_2}&= d_{11}\delta_{R_0}-(d_4-\frac15 d_7)  \, 16 \frac{r^6}{R_0^8}\delta_{R_0} \nonumber\\& \quad + (d_4+\frac45 d_6 +\frac{3}{25} d_7)\, 20 \frac{r^2}{R_0^4}\delta_{R_0}\,.
\end{align}

\section{Fourier transformation of\\ contact interactions}
\label{Fouriertrafo}

In the following we give the Fourier transformation of the contact
contributions. The LO contacts are momentum independent and their
Fourier transformation is given by
\begin{align}
&\quad  \int \!\!\! \frac{d^3 q}{(2\pi)^3} \, V_{\text{cont}}^{\text{LO}} f_{\text{local}}(q^2) e^{i \bf{q}\cdot \bf{r}} \\
&= V_{\text{cont}}^{\text{LO}} \, \int \!\!\! \frac{d^3 q}{(2\pi)^3}  f_{\text{local}}(q^2) e^{i \bf{q}\cdot \bf{r}}= V_{\text{cont}}^{\text{LO}} \, \delta_{R_0}({\bf r})\,, \nonumber
\end{align}
where $f_{\text{local}}(q^2)$ is a local momentum space regulator.

The first four NLO contact interactions are proportional to $q^2$ and
contain spin and isospin operators which are not dotted into momentum
operators. Writing the $q^2$ dependence explicitly, the Fourier
transformation is given by
\begin{align}
&\quad \int \!\!\! \frac{d^3 q}{(2\pi)^3} \, V_{\text{cont}}^{\text{NLO}} q^2 f_{\text{local}}(q^2) e^{i \bf{q}\cdot \bf{r}} \\
&= -V_{\text{cont}}^{\text{NLO}} \Delta \, \int \!\!\! \frac{d^3 q}{(2\pi)^3} f_{\text{local}}(q^2) e^{i \bf{q}\cdot \bf{r}}= -V_{\text{cont}}^{\text{NLO}} \Delta \, \delta_{R_0}({\bf r})\,. \nonumber
\end{align}

To Fourier transform the spin-orbit interaction we employ the test
function $\psi$:
\begin{align}
&\quad \bra{\bf{r}} \widehat{O}_{\text{LS}}\ket{\psi} \\
&= \int \!\!\! \frac{d^3 p}{(2\pi)^3}  \frac{d^3 p'}{(2\pi)^3} d^3r'\left\langle \bf{r}|\bf{p}'\right\rangle \bra{\bf{p}'}\widehat{O}_{\text{LS}}\ket{\bf{p}} \left\langle \bf{p}|\bf{r}'\right\rangle \left\langle \bf{r}'|\psi \right\rangle \nonumber \\
&= \int \!\!\! \frac{d^3 p}{(2\pi)^3}  \frac{d^3 p'}{(2\pi)^3} d^3r'e^{i \bf{p}' \cdot \bf{r}}  e^{-i \bf{p} \cdot \bf{r}'} \bra{\bf{p}'}\widehat{O}_{\text{LS}}\ket{\bf{p}} \psi(\bf{r}') \nonumber \\
&= \frac{C_5}{2}\int \!\!\! \frac{d^3 q}{(2\pi)^3}  \frac{d^3 k}{(2\pi)^3} d^3r' i(\fet \sigma_1 + \fet \sigma_2)\cdot (\bf{q} \times {\bf k})\nonumber \\ & \quad \times e^{i \frac{\textbf{q}}{2}  \cdot (\bf{r}+\bf{r}')}  e^{i{\bf k} \cdot (\bf{r} -\bf{r}')} \psi({\bf{r}}') f_{\text{local}}(q^2) \nonumber\\
&= \frac{C_5}{2}\int \!\!\! \frac{d^3 q}{(2\pi)^3}  \frac{d^3 k}{(2\pi)^3} d^3r' i\epsilon^{\alpha \beta \gamma}(\fet \sigma_1 + \fet \sigma_2)_{\alpha} q_{\beta} \nonumber \\ & \quad \times e^{i \frac{\textbf{q}}{2}  \cdot (\bf{r}+\bf{r}')} (i\partial_{\gamma}' e^{i{\bf k} \cdot (\bf{r} -\bf{r}')}) \psi({\bf{r}'}) f_{\text{local}}(q^2) \nonumber\\
&= -\frac{C_5}{2}\int \!\!\! \frac{d^3 q}{(2\pi)^3}  \frac{d^3 k}{(2\pi)^3} d^3r' i\epsilon^{\alpha \beta \gamma}(\fet \sigma_1 + \fet \sigma_2)_{\alpha} q_{\beta} \nonumber \\ & \quad \times (i\partial_{\gamma}'e^{i \frac{\textbf{q}}{2}  \cdot \bf{r}'}\psi({\bf{r}'})) e^{i{\bf k} \cdot ({\bf{r}} -{\bf{r}'})}  f_{\text{local}}(q^2) e^{i \frac{\textbf{q}}{2}  \cdot \bf{r}} \nonumber\\
&= \frac{C_5}{4}\int \!\!\! \frac{d^3 q}{(2\pi)^3} i\epsilon^{\alpha \beta \gamma}(\fet \sigma_1 + \fet \sigma_2)_{\alpha} q_{\beta} q_{\gamma}\psi({\bf{r}})  f_{\text{local}}(q^2)e^{i \bf{q} \cdot \bf{r}}  \nonumber \\& \quad -\frac{C_5}{2}\int \!\!\! \frac{d^3 q}{(2\pi)^3}i \epsilon^{\alpha \beta \gamma}(\fet \sigma_1 + \fet \sigma_2)_{\alpha} q_{\beta} (i\partial_{\gamma}\psi({\bf{r}}))  f_{\text{local}}(q^2)e^{i \bf{q}  \cdot \bf{r}}  \nonumber\\
&= -\frac{C_5}{2}  \epsilon^{\alpha \beta \gamma}(\fet \sigma_1 + \fet \sigma_2)_{\alpha} \partial_{\beta}\left(\int \!\!\! \frac{d^3 q}{(2\pi)^3}   f_{\text{local}}(q^2)e^{i \bf{q}  \cdot \bf{r}}\right) (i\partial_{\gamma}\psi(\bf{r}))  \nonumber\\
&= -\frac{C_5}{2} \frac{\partial_r\delta_{R_0}}{r} \epsilon^{\alpha \beta \gamma}(\fet \sigma_1 + \fet \sigma_2)_{\alpha}  {\bf{r}}_{\beta}(i\partial_{\gamma}\psi(\bf{r}))  \nonumber\\
&= -\frac{C_5}{2} \frac{\partial_r \delta_{R_0}}{r}{\bf S} \cdot  i {\bf{r}} \times \nabla \psi({\bf{r}}) = \frac{C_5}{2} \frac{\partial_r\delta_{R_0}}{r}  \, \bf{L} \cdot \bf{S}\, \psi(\bf{r})\,.  \nonumber
\end{align}
Here we used partial integration and the antisymmetry of $ \epsilon^{\alpha \beta \gamma}$ 
in line 5 and 6, respectively, and ${\bf{L}}=-i\bf{r} \times \nabla$ in the last line. 

The Fourier transformation of the tensorial contact operators is given
by
\begin{align}
&\quad \int \!\!\! \frac{d^3 q}{(2\pi)^3} \, V_{\text{cont}}^{\text{tens}} f_{\text{local}}(q^2) \fet \sigma_1 \cdot {\bf{q}} \, \fet \sigma_2 \cdot {\bf{q}} \,e^{i \bf{q}\cdot \bf{r}} \\
&= -V_{\text{cont}}^{\text{tens}} \sigma_1^i \sigma_2^j \partial^i \partial^j \int \!\!\! \frac{d^3 q}{(2\pi)^3} \,  f_{\text{local}}(q^2) \,e^{i \bf{q}\cdot \bf{r}} \nonumber\\
&= -V_{\text{cont}}^{\text{tens}} \sigma_1^i \sigma_2^j \partial^i \partial^j \delta_{R_0}(\bf{r}) \nonumber \\
&=-V_{\text{cont}}^{\text{tens}} \sigma_1^i \sigma_2^j \partial^i \left(\frac{x^j}{r} \partial_r\delta_{R_0}(\bf{r})\right) \nonumber \\
&= V_{\text{cont}}^{\text{tens}} \left[\fet \sigma_1 \cdot \hat{\bf{r}}\,\fet \sigma_2 \cdot \hat{\bf{r}}\left( \frac{\partial_r\delta_{R_0}(\bf{r})}{r}- \partial^2_r\delta_{R_0}(\bf{r})\right) \right. \nonumber \\&\quad -\left.\fet \sigma_1 \cdot \fet \sigma_2 \frac{\partial_r\delta_{R_0}(\bf{r})}{r} \right] \nonumber\,.
\end{align}

\end{document}